\documentclass[onecolumn,showpacs,nofootinbib,aps,superscriptaddress,prd,11pt]{revtex4}
\usepackage{graphicx,dcolumn,bm,amsfonts,amsmath,color,xcolor,amsthm}
\usepackage[colorlinks=true, citecolor=blue, linkcolor=blue,urlcolor=blue]{hyperref}
\usepackage{slashed}
\usepackage{ulem}

\newcommand\add[1]{{\color{blue}#1}}

\allowdisplaybreaks

\begin{document}
\title{Searching the possibility of $a_0(1450)$ scalar state being a diquark structure via charmed meson semileptonic decays}

\author{Ya-Lin Song}
\author{Yin-Long Yang\footnote{Ya-Lin Song and Yin-Long Yang contributed equally to this work.}
}
\affiliation{Department of Physics, Guizhou Minzu University, Guiyang 550025, P.R.China}
\author{Ye Cao}
\affiliation{Southern Center for Nuclear-Science Theory (SCNT), Institute of Modern Physics, Chinese Academy of Sciences, Huizhou 516000, China}
\author{Xue Zheng}
\affiliation{Department of Physics, Guizhou Minzu University, Guiyang 550025, P.R.China}
\author{Hai-Bing Fu}
\email{fuhb@gzmu.edu.cn}
\affiliation{Department of Physics, Guizhou Minzu University, Guiyang 550025, P.R.China}
\address{Institute of High Energy Physics, Chinese Academy of Sciences, Beijing 100049, P.R.China}

\begin{abstract}
The internal structure of light scalar state $a_0(1450)$ has not been definitively determined, it may consist of multiple possible states. Among them, it has the possibility of being regarded as a diquark state. Based on this possibility, we use QCD light-cone sum rules to study the semileptonic decay process $D \to a_0(1450)\ell \nu_\ell $ with $\ell=(e, \mu)$ to verify its rationality. Firstly, we construct two types of twist-2 light-cone distribution amplitude schemes based on the light-cone harmonic oscillator model, and present their moments $\langle\xi^{n}\rangle |_{\mu}$ and Gegenbauer moments $a_{n}(\mu)$ at $\mu_0=1~{\rm GeV}$ and $\mu_k= 1.4~{\rm GeV}$ for $n=(1,3,5)$. In the large recoil region, we obtain the transition form factors (TFFs): $f_+^{\rm (S1)}(0) = 0.836_{-0.116}^{+0.119}$, $f_+^{\rm (S2)}(0)=0.767_{-0.105}^{+0.106}$ and $f_-(0)=0.630_{-0.077}^{+0.078}$. A simplified series expansion $z(q^2, t)$ is used to extrapolate TFFs to the entire physical $q^2$-region. For $q^2=10^{-5} ~{\rm GeV}^2$, we compute angular distribution of the differential decay width ${d\Gamma}/{d\cos\theta_\ell }$ over the range $\cos\theta_\ell \in [-1,1]$. Subsequently, we obtain differential decay widths and branching fractions for $D^0 \to a_0(1450)^- \ell^+ \nu_\ell $ and $D^- \to a_0(1450)^0 \ell^- \bar{\nu}_\ell $, where the branching fractions being of order $10^{-6}$. Finally, we analyze three angular observables for the semileptonic decay process $D^- \to a_0(1450)^0 \ell^- \bar{\nu}_\ell $, the forward-backward asymmetry ${\cal A}_{\rm FB}$, lepton polarization asymmetry ${\cal A}_{\lambda_\ell}$ and $q^2$-differential flat term~${\cal F}_{\rm H}$.
\end{abstract}
\maketitle

\section{Introduction}\label {Sec:I}
Over the past few decades, with the steady improvement in measurement precision of hadronic decay process by BESIII, LHCb and Belle II, the study of light diquark scalar state decay properties has gradually become an important focus in hadron physics. As a typical isovector light scalar state above $1~{\rm GeV}$,  $a_0(1450)$ (with $J^{PC}=0^{++},~I=1$) has long been controversial regarding its fundamental structure. Up to now, many possible explanations for its internal components, including traditional ground $q \bar{q}$ states~\cite{Cheng:2005nb, Rui:2018mxc, Chai:2021pyp, Guo:2022xqu, Han:2013zg}, tetraquark states $(qq\bar{q} \bar{q})$~\cite{Jaffe:1976ig, Weinstein:1983gd} and hybrid states~\cite{Klempt:2021nuf}. This makes $a_0(1450)$ one of the challenging topics in nonperturbative quantum chromodynamics (QCD). Based on the accumulating experimental data, two widely discussed classification scenarios for scalar states have been proposed: In the first picture (P1), light scalar states such as $f_0(980), ~ a_0(980)$, and $K_0^*(700)$ are regarded as ground $q\bar{q}$ states, while the nonet states around $1.5 ~\rm{GeV}$ $(\mathrm{such~as} ~a_0(1450)$, and $K_0^*(1430))$ are interpreted as their first radial excited states. In second picture (P2), the states such as $f_0(1370), ~ a_0(1450)$, and $K_0^*(1430)$ are treated as lowest-lying $P$-wave $q\bar{q}$ states, whereas the nonet states below $1~\rm{GeV}$ are viewed as tetraquark bound states~\cite{Brito:2004tv, Klempt:2007cp}.

Currently, more experimental and theoretical studies tend to support the P2 scenario. For scalar states below $1~{\rm GeV}$, experiments (e.g., BESIII~\cite{BESIII:2018sjg, BESIII:2021drk, BESIII:2023opt, BESIII:2023wgr} and CLEO~\cite{CLEO:2009ugx}) suggest that their internal structure are not simple ground $q\bar{q}$ states based on measured data. Instead, these states are more likely to be described as a tetraquark state or a mixed state. Moreover, the review ``Scalar Mesons below $1~{\rm GeV}$'' provided by the particle data group (PDG) points out that scalar state ($f_0(500)/\sigma$, $K^*_0(700)/\kappa$, $f_0(980)$, $a_0(980)$) below $1~{\rm GeV}$ are more likely a nonet dominated by tetraquark components, rather than traditional ground $q\bar{q}$ state~\cite{ParticleDataGroup:2020ssz}. As for the scalar state $a_0(1450)$ above $1~{\rm GeV}$, it is regarded as the lowest-lying $P$-wave $q\bar{q}$ state in P2 scenario, this view has been supported by many studies. For example, in $p\bar{p}$ annihilation experiments, Crystal Barrel collaboration made the first measurements of branching fractions for $a_0(1450) \to \pi \eta, K\bar{K}$, $\pi \eta^{\prime}$. The results are consistent with the expectations from flavor SU(3) symmetry for conventional ground $q\bar{q}$ states, thereby classifying $a_0(1450)$ as a member of the scalar state nonet~\cite{CrystalBarrel:1994arw}. Lattice QCD calculations show that the mass of $a_0(1450)$ is about $1.42~{\rm GeV}$. The result is obtained by analyzing the correlation function of the $\psi \bar\psi$ interpolating field and subtracting the contribution from the $\eta^{\prime} \pi$ ghost state, which is consistent with expectations for conventional ground $q\bar{q}$ states~\cite{Mathur:2006bs}. In addition $a_0(1450)$ resonance is explicitly regarded as a conventional low-lying $P$-wave $q\bar{q}$ state within framework of the naive quark model. Based on this, the authors further investigated decay chain $J\psi \to \gamma\eta_c \to \gamma\pi^0 a_0(1450) \to \gamma\pi^0 a_0(980)f_0(500)$ and obtained a branching ratio of the order of $10^{-6}$. This result is consistent with the expectations of lattice QCD calculations~\cite{Cheng:2020qzc, Lee:1999kv}. It is worth noting that the latest review by PDG explicitly notes that scalar states below $1~\rm{GeV}$ $(\rm{such ~as ~}$ $a_0(980))$ tend to exhibit tetraquark characteristics, while above $1~\rm{GeV}$ (such as $a_0(1450))$ are more closely consistent with conventional ground $q\bar{q}$ states~\cite{ParticleDataGroup:2024cfk}. This is consistent with the description of P2 scenario.

In addition, the University of Notre Dame and Argonne National Laboratory used the Argonne ZGS accelerator to study reaction $\pi^- p \to n K^0_S K^0_L$, and measured the width of $a_0(1450)$-state to be $\Gamma_{a_0(1450)}=79\pm{10}~\mathrm{MeV}$~\cite{Cason:1976fn}. OBELIX Collaboration performed a partial-wave analysis of $\bar{p}p$ annihilations at rest into $K^0_S K^+ \pi^-$ and $K^0_S K^- \pi^+$ final states, obtaining a width of $\Gamma_{a_0(1450)}=80\pm5~\mathrm{MeV}$~\cite{OBELIX:1998sco}. Recently in 2019, Belle Collaboration analyzed two-photon fusion process $\gamma\gamma \to \eta\pi^0$ and observed a scalar resonance consistent with $a_0(1450)$ in this process. After fitting results, the decay width is obtained to be $\Gamma_{a_0(1450)}=65.0^{+2.1 ~+99.1}_{-5.4 ~-32.6}~{\rm MeV}$~\cite{Belle:2009xpa}. Based on the above discussion, both experimental and theoretical results provide the possibility to classify it into the P2 scenario. Therefore, the light scalar $a_0(1450)$-state can be regarded as conventional ground $q\bar{q}$ state. In this context, it is meaningful to probe the internal structure of light scalar $a_0(1450)$-state via semileptonic decay process $D \to a_0(1450)\ell \nu_\ell $, and this is one of the important motivation for this work.

As the lightest particle containing a charm quark ($c$-quark), semileptonic decays of $D$-meson have a simpler decay mechanisms and final state interactions, which can provide an ideal platform for investigating the properties of meson or diquark state. On the experimental side, the BESIII ~\cite{Liu:2019tsi, Zhang:2019tcs, Yang:2018qdx, BESIII:2016gbw, BESIII:2021mfl, BESIII:2015tql, BESIII:2021pvy, BESIII:2015kin}, BaBar~\cite{BaBar:2014xzf}, Belle~\cite{Belle:2006idb}, and CLEO collaborations~\cite{CLEO:2011ab, CLEO:2004arv, CLEO:2009dyb, CLEO:2009svp, CLEO:2005rxg} have measured some semileptonic decay processes involving $D$-mesons. Among them, BESIII collaboration made the first experimental observation of a semileptonic decay process involving light scalar state $a_0(980)$. Utilizing an $e^+ e^-$ collision collected at a center-of-mass energy of $3.773~\rm{GeV}$, corresponding to an integrated luminosity of $2.93~\rm{fb^{-1}}$, BESIII reported the observations of decays $D^0 \to a_0(980)^{-} e^{+} \nu_{e}$ and $D^+ \to a_0(980)^{0} e^+ \nu_{e}$. The measured absolute branching fractions are on the order of $10^{-4}$, with significance of these observations reached $6.4~\sigma$ and $2.9~\sigma$, respectively~\cite{BESIII:2016gbw}. Compared with the increasing experimental and theoretical studies on light scalar state such as $a_0(980)$, experimental data for the heavier $a_0(1450)$ are remain scarce, and semileptonic decay process $D \to a_0(1450)\ell \nu_\ell $ has not been measured experimentally. This lack of data limits our understanding of the internal structure of light scalar states and their classification within the hadronic spectrum. In this regard, the semileptonic decay process $D \to a_0(1450)\ell \nu_\ell $ with the $a_0(1450)$ in the final state can provide valuable insights into the intrinsic properties of scalar states.

Theoretically, the heavy to light transition form factors (TFFs) in semileptonic decay processes are conventionally computed using nonperturbative approaches. Currently, several theoretical methods have been developed to study the $D \to a_0(1450)$ TFFs. For instance, in 2011, R. C. Verma employed the covariant light-front quark model (CLFQM) to investigate $a_0(1450)$ decay constant and $D \to a_0(1450)$ TFFs~\cite{Verma:2011yw}. Additionally, both relativistic quark model (RQM) and LCSR approachs are matching the P2 scenario and calculated $D \to a_0(1450)$ TFFs $f_{\pm} (q^2)$ ~\cite{Galkin:2025emi,Huang:2021owr}. However, these two theoretical frameworks differ. RQM calculates the quark-antiquark bound state wave function using a quasipotential approach and takes relativistic corrections into account. In contrast, the latter focuses on twist-2 light-cone distribution amplitudes (LCDAs) and uses Gegenbauer polynomial expansion to describe its behavior. Developed in 1980s. LCSR approach effectively combines the SVZ sum rules (SVZSR) with theory of hard exclusive processes~\cite{Balitsky:1989ry, Chernyak:1990ag}. It is widely regarded as an advanced and well-established tool for handling heavy-to-light transitions~\cite{Cheng:2017bzz, Tian:2023vbh, Gao:2019lta, Duplancic:2008ix}. Compared with traditional SVZ sum rules, the LCSR approach can parameterize nonperturbative contributions in terms of LCDAs, which allows for the quantification of higher-twist nonperturbative corrections. This method only involves a single Borel transformation and dispersion relations, optimizing computational steps. Consequently, the LCSR approach is adopted in this work to investigate semileptonic decay process $D \to a_0(1450) \ell \nu_\ell $. Furthermore, compared with previous LCSR analysis~\cite{Huang:2021owr}, the TFF in our work account for contributions from both $a_0(1450)$-state twist-2 and twist-3 LCDAs.

As an important nonperturbative parameter in the transition process $D \to a_0(1450)$, the $a_0(1450)$-state twist-2 LCDA plays an key role in determining behavior and precision of TFFs, and it incorporates long-range QCD dynamics at low energy scales. Therefore, one have also paid great attention to determining the precise behavior of its twist-2 LCDA. Conventionally, the $a_0(1450)$-state twist-2 LCDA can be expanded in terms of Gegenbauer series, and a truncated form retaining only the first few terms is commonly adopted~\cite{LatticeParton:2022zqc}. These coefficients of Gegenbauer series (also called Gegenbauer moments) can be calculated using the QCD sum rules method. Furthermore, to provide a new phenomenological perspective on the internal dynamics information of $a_0(1450)$-state, we can also construct its twist-2 LCDA through light-cone harmonic oscillator (LCHO) model. This model is based on brodsky-huang-lepage (BHL) prescription and incorporates Wigner-Melosh rotations, which together constitute theoretical basis of the framework. Meanwhile, it connects equal-time wave functions in the rest frame to light-cone wave functions in the infinite-momentum frame, thereby transforming them into a relativistic form expressed in light-cone coordinates. LCHO model can also describe the spatial component and spin component of wave function at the same time, which provides an effective characterization for the momentum distribution internal the meson~\cite{Huang:1994dy}. This makes it suitable for applications to $D \to a_0(1450)$ TFFs. Furthermore, we adopt two distinct longitudinal correction functions $\varphi^{\rm (S1, S2)}_{a_0(1450)}(x)$ in twist-2 wavefunction to form two twist-2 LCDA schemes for $a_0(1450)$-state. By comparing observables of the semileptonic decay process $D \to a_0(1450)\ell \nu_\ell $ calculated under two schemes, we can not only test standard model (SM) but also help examine the reliability and feasibility of our LCHO model.

The remaining parts of this paper are organized as follow. Section~\ref{sec:II} presents the expressions for $D \to a_0(1450)$ within the LCSR framework. Meanwhile, based on the LCHO model, we construct two twist-2 LCDA schemes of $a_0(1450)$, and calculate the moments $\langle\xi^{n}\rangle |_{\mu}$ and the Gegenbauer moments $a_{n }(\mu)$. In Section~\ref{sec:III}, we present numerical analyses and discussions, including TFFs, angular distributions, decay widths, decay branching fractions and three angular observables. In Section~\ref{Sec:IV}, we conclude this paper with a brief summary.

\section{Theoretical Framework}\label {sec:II}
\begin{figure}[t]
\begin{center}
\includegraphics[width=0.45\textwidth]{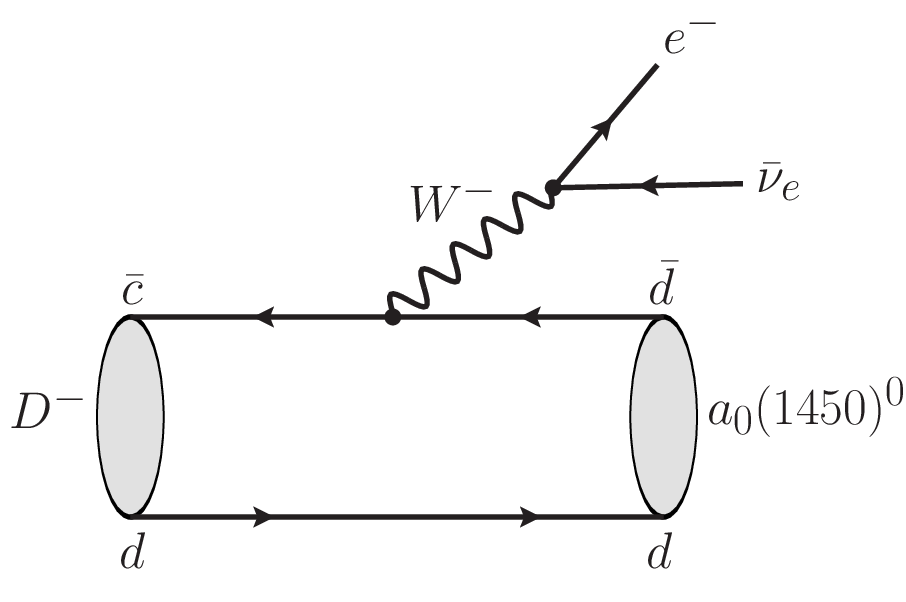}
\end{center}
\caption{Feynman diagram representing the tree-level charged current process $D^- \to a_0(1450)^0 \ell^- \bar{\nu}_\ell $.}
\label{Fig:feynman}
\end{figure}
Based on the possibility for light scalar state $a_0(1450)$ can be regarded as a diquark $q\bar{q}$ state, it can exists in three charge configurations with the following quark compositions respectively: $a_0(1450)^+={u\bar{d}}$, $a_0(1450)^-={d\bar{u}}$ and $a_0(1450)^0= (u\bar{u}-d\bar{d})/\sqrt{2}$. The differences among them arise from distinct quark-antiquark flavor compositions of each state. Here we take the semileptonic decay $D^- \to a_0(1450)^0 \ell^- \bar{\nu}_\ell $ for instant. Its basic mechanism can be described through the feynman diagram is shown in Fig.~\ref{Fig:feynman}. In which, the anti-charm quark $\bar{c}$ transitions to an anti-down quark $\bar{d}$ via a virtual $W^-$ boson. Subsequently, $W^-$ decays into an $e^-$ and an $\bar{\nu}_{e}$, $d$-quark does not participate in the weak interaction and finally combines with produced $\bar{d}$ to form light scalar state $a_0(1450)^0$. This decay process can be described by an effective Hamiltonian $\mathcal{H}_{\text{eff}} = G_F|V_{cd}| \bar{c} \gamma_{\mu}(1-\gamma_{5})d \bar{\ell} \gamma^{\mu}(1-\gamma_{5}) \nu_\ell  /\sqrt{2} $. At the hadronic level, by sandwiching the free-quark amplitude between the initial and final meson states, one can obtain the decay amplitude for the semileptonic $D \to a_0(1450)\ell \nu_\ell $. This amplitude consists of a leptonic current and a hadronic current. Since the leptonic current does not participate in strong interactions, it can be calculated naturally. In contrast, the hadronic current involves non-perturbative effects that cannot be computed from first principles. It can usually be parameterized in terms of a set of Lorentz-invariant hadronic TFFs, $i.e.,$
\begin{align}
\langle a_0(1450)(p)&|\bar{c}\gamma_\mu\gamma_5d|\, D(p+q)\rangle  =-i\left[ f_+ (q^2)p_{\mu}+ f_- (q^2)q_{\mu} \right].
\end{align}
Thus, the complex hadron information has been transformed into a computable quantity. By integrating over the phase space, double differential decay width changed with squared momentum transfer and angle between momentum of $a_0(1450)$ and lepton $\ell$ in center of mass frame for lepton pair that can be written as:
\begin{align}
\frac{d^2\Gamma({D \to a_0(1450) \ell \nu{_\ell}})}{dq^2 d\cos\theta_\ell}
&=\frac{G_F^2 |V_{cd}|^2 m_D^3 \lambda^{1/2}}{256 \pi^3 c_{a_0(1450)}^2} \bigg(1-\frac{m_\ell ^2}{q^2} \bigg)^2 \bigg\{\lambda |f_+ (q^2)|^2 \bigg[1-\bigg(1-\frac{m_\ell ^2}{q^2} \bigg) \mathrm{cos}^2\theta_\ell  \bigg]
\nonumber \\
& +\bigg(1-\frac{m_{a_0(1450)}^2}{m_D^2} \bigg)^2  \frac{m_\ell ^2}{q^2}  \bigg[ |f_0 (q^2)|^2 +2\lambda^{1/2} \mathrm{Re} [ f_+ (q^2) f_0^*(q^2)]\bigg]\cos\theta_\ell\bigg\},
\label{dGammadcos}
\end{align}
where $G_F = 1.1663787(6)\times 10^{-5} \rm{GeV}^{-2}$ is Fermi coupling constant, $m_\ell$ is lepton mass and  $\lambda\equiv\lambda(1,m_{a_0(1450)}^2/m_D^2,q^2/m_D^2)$ with $\lambda(a,b,c)\equiv{a^2+b^2+c^2-2(ab+ac+bc)}$. Meanwhile the scalar TFF have the relationship with the traditional vector TFFs {\it i.e.} $f_0 (q^2)=f_+ (q^2)+{q^2}/(m_D^2-m_{a_0(1450)}^2) f_- (q^2)$. Specifically, the isospin factor $c_{a_0(1450)} = \sqrt{2}$ corresponds to $D^- \to a_0(1450)^0 \ell^- \bar{\nu}_\ell $, while $c_{a_0(1450)} = 1$ corresponds to $D^0 \to a_0(1450)^- \ell^+ \bar{\nu}_\ell $ based on the difference in electric charge. Generally, the differential decay width serves as a important bridge connecting theory and experiment, which is also considered as one of the effective means for testing SM. On the other hand, one can also focus on the angular distribution of differential decay width, along with three angular observables that are highly sensitive to new physics, $i.e.,$ the forward-backward asymmetry ${\cal A}_{\rm FB}(q^2)$, lepton polarization asymmetry ${\cal A}_{\lambda_\ell}$ and $q^2$-differential flat term ${\cal F}_{\rm H}(q^2)$. For the brevity of this short essay, there specific expressions are not present here, which can be found in Ref.~\cite{Cui:2022zwm}.

In order to derive the $D \to a_0(1450)$ TFFs based on the QCD LCSR, one can start from the two-point correlation function from vacuum to final state light $a_0(1450)$ state
\begin{align}
\Pi_{\mu}(p,q) = i\int d^4x e^{iq\cdot x} \langle a_0(1450)(p)|T\{J_n(x), j_n^{\dagger}(0)\}|0 \rangle,
\label{correlation}
\end{align}
where the current operators $J_n(x)=\bar{q}_2(x) \gamma_{\mu} \gamma_{5} c(x)$ describes the weak transition of $c\to q_2$ and $j_{n}^{\dagger}(0)=\bar{c} i\gamma_{5} q_1(0)$ represents the $D$-meson decay channel. The $q_1$ and $q_2$ represent light quarks. This correlation function can be processed in different kinematic regions, allowing for a reasonable matching between its phenomenological representation and its theoretical expression. The specific procedure is as follows: Firstly, in the time-like $q^2$ region, we insert a complete set of intermediate states with $D$-meson quantum numbers to obtain the hadronic representation,
\begin{align}
\Pi_{\mu}^{\text{had}}(p,q) &= \frac{\langle a_0(1450)|\bar{q}_2 \gamma_\mu \gamma_5 c|D\rangle\langle D|\bar{c}i\gamma_5 q|0\rangle}{m_D^2 - (p+q)^2} + \sum_{\rm H} \frac{\langle a_0(1450)|\bar{q}_2 \gamma_\mu \gamma_5 c|D^{\rm H}\rangle\langle D^{\rm H}|\bar{c}i\gamma_5 q_1|0\rangle}{m_{D^{\text{H}}}^2 - (p+q)^2},
\end{align}
where vacuum-to-meson matrix element can be defined as $\langle D|\bar{c}i\gamma_5 q_1 |0\rangle =m^2_D f_D/m_c$. With this definition, we can further derive
\begin{align}
\Pi_{\mu}(p,q) &= \frac{m_{D}^2 f_{D}}{m_c \left[m_{D}^2 - (p+q)^2\right]} \left[f_+(q^2)p_{\mu} + f_-(q^2)q_{\mu}\right] + \int_{s_0 }^{\infty} ds \frac{\rho_+^{\rm H}(s)p_{\mu} + \rho_-^{\rm H}(s)q_{\mu}}{s - (p+q)^2}.
\end{align}
Here, the contribution from ground state is isolated. Contributions from higher resonances and continuum states are replaced by a spectral density function $\rho_{+,-}^{\rm H}(s)$. Due to the complexity of multi-hadron continuum states at high energies, it is difficult to obtain an explicit analytic expression for this function. To address this, we invoke quark-hadron duality to relate it to the corresponding QCD representation.
\begin{align}
\rho^{\rm H}_{+,-}(s) = \rho^{\rm{QCD}}_{+,-}(s)\theta(s - s_0).
\end{align}
Additionally, the correlation function can be calculated in the deep Euclidean region $q^2=-Q^2\ll 0$ with operator product expansion (OPE) near the light-cone ($x^2 \approx 0$). Then, the QCD expression can be derived by contracting heavy $c$ quark to a full quark propagator
\begin{align}
\langle 0|c_{\alpha}^{j}(x)\bar{c}_{\beta}^{j}(0)|0\rangle
&= i \int \frac{d^4k}{(2\pi)^4} e^{-ik\cdot x} \left[ \delta^{ij} \frac{\slashed{k} + m_c}{k^2 - m_c^2} + \cdots \right]_{\alpha\beta}.
\end{align}
The first term corresponds to the free quark propagator, which provides the leading contribution. The second term arises from the one gluon contribution, which generally does not play a significant role in the sum rules for TFFs and can be safely neglected.
Currently, the scalar diquark state such as $a_0(1450)$ twist-4 LCDAs are still not well established. So we retain its twist-2 and twist-3 LCDAs in this paper which are widely studied with some other approaches.
After substituting free quark propagator into correlation function, one can obtain the OPE results. By applying Borel transformation and employing dispersion relations to match QCD representation with the hadronic representation, the finial analytical expressions for TFFs within LCSR framework can be written as
\begin{align}
&f_{+}^{\rm (S1,S2)} (q^2) = \frac{m_c ~ \bar{f}_{a_0(1450)}}{m_D^2 f_D} ~ \int_{u_0}^1 ~ du ~ e^{(m_{D}^2 -s(u))/M^2} ~ \bigg\{-\frac{m_c}{u} ~ \phi^{\rm (S1,S2)}_{2;a_0(1450)}(u,\mu) \,+\, m_{a_0(1450)}
\nonumber \\
&\qquad \qquad \,\,\,\, \times \phi_{3;a_0(1450)}^p(u,\mu) +\frac{m_{a_0(1450)}}{6} \bigg[ \frac{2}{u} \phi_{3;a_0(1450)}^\sigma(u,\mu)-\bigg(\frac{m_c^2-u^2 m_{a_0(1450)}^2 +q^2}{m_c^2 +u^2 m_{a_0(1450)}^2 -q^2}
\nonumber \\
&\qquad \qquad \,\,\,\, \times  \frac{d\phi_{3;a_0(1450)}^\sigma(u,\mu)}{du} - \frac{4u m_c^2 m_{a_0(1450)}^2}{(m_c^2 +u^2 m_{a_0(1450)}^2 -q^2)^2}  \phi_{3;a_0(1450)}^\sigma(u,\mu) \bigg) \bigg] \bigg\},
\\
&f_{-} (q^2) = \frac{ \bar{f}_{a_0(1450)}}{m_D^2 f_D} \int_{u_0}^1  du e^{(m_{D}^2 - s(u))/M^2} \bigg[ \frac{m_c}{u} {\phi_{3;a_0(1450)}^p (u,\mu)} +\frac{m_c}{6u} \frac{d\phi_{3;a_0(1450)}^\sigma(u,\mu)}{du} \bigg],
\label{TFFs}
\end{align}
where the lower limit of integration is $u_0 = \{ [(s-q^2-m_{a_0(1450)}^2)^2 +4m_{a_0(1450)}^2(m_{c}^2-q^2)]^{1/2}-(s-q^2-m_{a_0(1450)}^2) \}/(2m_{a_0(1450)}^2)$, $s(u) = (m_c^2 +u\bar{u}m_{a_0(1450)}^2 -\bar{u}q^2)/u$ with $\bar{u}=(1-u)$, $m_D$ and $f_D$ are the mass and decay constant of $D$-meson, $m_c$ is mass of the $c$-quark, and $s_0$ represents the continuum threshold parameter, $M^2$ is the Borel window.

As the main source of nonperturbative uncertainty in LCSR expressions, the twist-2 LCDA of $a_0(1450)$ is a universal nonperturbative quantity, it is appropriate to study it using nonperturbative QCD methods. Generally, the exploration of $\phi_{2;a_0}(x,\mu)$ can be carried out by combining nonperturbative QCD with phenomenological models. Thus, with the $\pi$-meson wave function as a reference and brodsky-huang-lepage (BHL) hypothesis as starting point, we establish a specific correspondence between the equal-time wave function in the rest frame and light-cone wave function~\cite{Wu:2010zc, Wu:2011gf}, which can be expressed as follows:
\begin{align}
\Psi_{2; a_0(1450)}(x,\mathbf{k}_{\perp}) = \sum_{\lambda_1\lambda_2} \chi_{a_0(1450)}^{\lambda_1\lambda_2}(x,\mathbf{k}_{\perp})\psi_{a_0(1450)}^R(x,\mathbf{k}_{\perp}).
\end{align}
For the form of spin wave function $\chi_{a_0(1450)}^{\lambda_1\lambda_2}(x,\mathbf{k}_{\perp})$, light meson wave function is usually transformed into the light-cone form to obtain complete spin wave function in the instantaneous SU(6) quark model~\cite{Huang:1994dy,Ma:1993ht,Wu:2007rt}. Starting from the instant-form, we take scalar state $a_0(1450)$ with spin $S=1$, orbital angular momentum $L=1$, and total angular momentum $J=0$. In the rest frame $(\boldsymbol{q}_1 +\boldsymbol{q}_2 =0)$, the spin wave function in instant-form (T) can be obtained,
\begin{align}
\chi_{a_0(1450)}^T=\frac{1}{\sqrt{2}}(\chi_1^\uparrow\chi_2^\downarrow-
\chi_2^\uparrow\chi_1^\downarrow),
\end{align}
where $\chi_{1/2}^{\uparrow/\downarrow}$ is the Pauli spinor of the triplet state, and the four-momenta of the two quarks are respectively: $q^{\mu}_1 =(q^0,\boldsymbol{q})$, $q^{\mu}_2 =(q^0,-\boldsymbol{q})$, $q^0=\sqrt{m^2 +\boldsymbol{q}^2}$. The instant-form $\vert J, s \rangle_T$ and light-cone form $\vert J, \lambda \rangle_F$ are related by Wigner rotation. For hadronic states with total angular momentum $J=0$, this rotation reduces to the identity matrix, expressed as $|J,\lambda \rangle_F=\sum_s U_{s \lambda}^J |J,s \rangle_T$. For a quark with spin-1/2, the corresponding Melosh transformation is given as follows:
\begin{align}
\chi^\uparrow(T)=\omega[(q^{+}+m)\chi^\uparrow(F)-q_R\chi^\downarrow(F)],
\nonumber \\
\chi^\downarrow(T)=\omega[(q^{+}+m)\chi^\downarrow(F)-q_L\chi^\uparrow(F)],
\end{align}
where $\omega=[2q^+(q^0+m)]^{-1/2}, \quad q_{R/L}=q_1\pm iq_2, \quad q^+=q^0+q^3$. Then, the spin wave function of $a_0(1450)$ state can be obtained,
\begin{align}
\chi_{a_0(1450)}(x,\mathbf{k}_\perp)=\sum_{\lambda_1,\lambda_2}C_0^F(x,\mathbf{k}_\perp,
\lambda_1,\lambda_2)\chi_{1}^{\lambda_1}(F)\chi_{2}^{\lambda_2}(F).
\end{align}
When expressed in terms of the instant-form momentum $q^\mu=(q^0,\boldsymbol{q})$, the component coefficient $C_{0}^F(x,\mathbf{k}_\perp,\lambda_1,\lambda_2)$ can be calculated, and its specific form is as follows:
\begin{align}
C_0^F(x,\mathbf{k}_\perp,\uparrow,\downarrow)&= + \frac{m}{\sqrt{2(m^2+\mathbf{k}_{\perp}^2)}}, \nonumber \\
C_0^F(x,\mathbf{k}_\perp,\downarrow,\uparrow)&=-\frac{m}{\sqrt{2\big(m^2+\mathbf{k}_{\perp}^2)}}, \nonumber \\
C_0^F(x,\mathbf{k}_\perp,\uparrow,\uparrow)&=- \frac{(k_1-ik_2)}{\sqrt{2(m ^2+\mathbf{k}_\perp^2)}}, \nonumber \\
C_0^F(x,\mathbf{k}_\perp,\downarrow,\downarrow)&=-\frac{(k_1+ik_2)}{\sqrt{2(m^2+\mathbf{k}_\perp^2)}},
\end{align}
these coefficients satisfy the following normalization relation: $\sum_{\lambda_{1},\lambda_{2}}C_{0}^{F}(x,\mathbf{k}_{\perp},\lambda_{1},\lambda_{2})^*
C_{0}^{F}(x,\mathbf{k}_{\perp},\lambda_{1},\lambda_{2})=1$. In addition, apart from the ordinary helicity component $(\lambda_1 +\lambda_2=0)$, there exist higher helicity components $(\lambda_1 +\lambda_2 = \pm1)$, while the instant-form wave function contains only the ordinary helicity component. Then, the spin wave function can be defined as
\begin{align}
\chi_{a_0(1450)}^{\lambda_1\lambda_2}(x,\mathbf{k}_{\perp})= \frac{\hat{m}_q^2}{\sqrt{\mathbf{k}_\perp^2 + \hat{m}_q^2}}.
\end{align}
Moreover, The BHL description establishes an equivalence between the spatial wave function $\Psi_{2;a_0(1450)}^R(x,\mathbf{k}_{\perp})$ and the equal-time wave function, thereby allowing us to avoid the complex problem of solving an infinite set of coupled integral equations when determining the specific form of the wave function.. This leads us to the following result:
\begin{align}
\psi_{2;a_0(1450)}^R(x,\mathbf{k}_{\perp}) = A \varphi_{a_0(1450)}(x) \exp \Big[ - \frac{\mathbf{k}_\perp^2 + \hat{m}_q^2}{8 \beta^2 {x}{\bar{x}}} \Big],
\end{align}
where the $\mathbf{k}_\perp$ denotes transverse momentum, $A$, $\hat{m}_q$ denote normalization constant and light quark mass, respectively~\cite{Cao:1997hw, Huang:2004su}. According to Ref.~\cite{Zhong:2021epq}, the harmonic-oscillator exponential factor $\exp\left[-(\mathbf{k}_\perp^2+m_q^2)/(8\beta^2x\bar{x})\right]$ primarily controls broadening in the transverse momentum $\mathbf{k}_\perp$. On the other hand, due to the consideration of the existence of quark spin, this will cause additional corrections to the longitudinal distribution of the distribution amplitude. For this, we can control the longitudinal distribution by introducing the function $\varphi_{a_0(1450)}(x)$, which allows adjustment of the $x$-direction shape (including endpoint behavior and width) of twist-2 LCDA for $a_0(1450)$. Usually, a form similar to the traditional Gegenbauer polynomial expansion is adopted. Based on this, we consider the first scheme:
\begin{align}
&\varphi^{\rm (S1)}_{a_0(1450)}(x)=C_1^{3/2}(x-\bar x),
\label{varphi1}
\end{align}
Since the twist-2 distribution amplitude of the scalar state $a_0(1450)$ is antisymmetric under $x\leftrightarrow (1-x)$, the zeroth Gegenbauer moment to zero. To more comprehensively validate the applicability of the model, we also consider the second scheme:
\begin{align}
\varphi^{\rm (S2)}_{a_0(1450)}(x)=(x\bar{x})^{\alpha}C_1^{3/2}(x-\bar x).
\label{varphi2}
\end{align}
This structure is optimized by introducing a factor $(x \bar{x})^{\alpha}$, this factor ensures that the expression approaches the theoretical limit $(x \bar{x})^{\alpha}=6x\bar{x}$ as $\mu \to \infty$~\cite{Lepage:1980fj}.

After integrating over the transverse momentum $\mathbf{k}_{\perp}$, the final twist-2 LCDA of the $a_0(1450)$ can be obtained,
\begin{align}
\phi^{\rm(S1, S2)}_{2;a_0(1450)}(x,\mu) &= \frac{A \hat{m}_q \beta}{4 \sqrt{2} \pi^{3/2}} \sqrt{x \bar{x}} \varphi^{\rm(S1, S2)}_{a_0(1450)}(x) \left\{\mathrm{Erf}\left[\sqrt{\frac{\hat{m}_q^2 + \mu^2}{8\beta^2 x\bar{x}}} \right]-\mathrm{Erf}\left[\sqrt{\frac{\hat{m}_q^2}{8\beta^2 x\bar{x}}}\right]\right\},
\end{align}
where $\mathrm{Erf}(x) = 2{\int_{0}^{x}e^{-t^2}dx}/{\sqrt{\pi}}$ is the error function, and we take $\hat{m}_q = 250~\rm{MeV}$. The model parameters of the above two amplitude schemes can be determined by the following criteria,
\begin{itemize}
  \item The average value of squared transverse momentum $\left\langle \textbf{k}_{\perp}^2\right\rangle$~\cite{Fu:2014uea}:
\begin{align}
\langle\mathbf{k}_\perp^2\rangle = \frac{\int dxd^2\mathbf{k}_\perp|\mathbf{k}_\perp|^2|\Psi_{2; a_0(1450)}^{\rm (S1,S2)}(x,\mathbf{k}_{\perp})|^2}{\int dxd^2\mathbf{k}_\perp|\Psi_{2; a_0(1450)}^{\rm (S1,S2)}(x,\mathbf{k}_{\perp})|^2}, \hspace{-0.65cm}
\end{align}
which is consistent with the choice of Ref.~\cite{Wu:2010zc} for light diquark states or light meson.
\end{itemize}

\begin{itemize}
  \item The Gegenbauer moments $a_n^{\rm (S1,S2)}(\mu)$ can be derived by the following way:~\cite{Zhong:2011rg}:
\begin{align}
a_n^{\rm (S1,S2)} (\mu)=\frac{\int_0^1dx\phi_{2;a_0(1450)}^{\rm (S1,S2)} (x,\mu)C_n^{3/2}(\xi)}{\int_0^1 dx 6x\bar{x}[C_n^{3/2}(\xi)]^2},
\label{gegenbauer}
\end{align}
where $\xi=(2x-1)$. Generally, the properties of twist-2 LCDA are mainly determined by its first few terms.
\end{itemize}

For the first scheme, we adopt the first-order Gegenbauer moment and average value of squared transverse momentum $\left\langle \textbf{k}_{\perp}^2\right\rangle$ to determine model parameters $A^{\rm (S1)}$ and $\beta^{\rm (S1)}$. For the model parameters $A^{\rm (S2)}$, $\beta^{\rm (S2)}$ and $\alpha^{\rm (S2)}$ in second
scheme, we adopt the $\left\langle \textbf{k}_{\perp}^2\right\rangle$, first and third\add{-}order Gegenbauer moments.
For the $a_0(1450)$ state's twist-2 LCDA $\phi_{2;a_0(1450)}^{\rm (S1,S2)}(x,\mu)$, the $n$th-order moment $\langle\xi^{n}\rangle|_\mu^{\rm (S1,S2)}$ -- an important quantity for the nonperturbative momentum distribution -- is defined as
\begin{align}
\langle\xi^{n}\rangle|_\mu^{\rm (S1,S2)}=\int_0^1 dx(2x-1)^{n}\phi_{2;a_0(1450)}^{\rm (S1,S2)}(x,\mu).
\label{moment}
\end{align}
To better compare LCHO model, we also adopt the truncated form of $a_0(1450)$-state distribution amplitude and substitute it into our TFFs to compute the subsequent physical observables~\cite{Cheng:2005nb},
\begin{align}
\phi_{2;a_0(1450)}^{\rm{TF}}(x,\mu)=6x\bar{x}\left[a_0(\mu)+\sum_{n=1}^{\mathcal{N}=3}a_n(\mu)
C_n^{3/2}(\xi)\right].
\label{TF}
\end{align}
Finally, the $a_0(1450)$-state twist-3 LCDAs can be generally expanded into a series of Gegenbauer polynomials and taken truncated form to remain the first few terms, specific form is as follows~\cite{Lu:2006fr}:
\begin{align}
\phi_{3;a_0(1450)}^{p}(x,\mu) &= 1+\sum_{n=1}^{\mathcal{N}=2} a_{n}^{p}(\mu) C_{n}^{1/2}(\xi), \nonumber \\
\phi_{3;a_0(1450)}^{\sigma}(x,\mu) &= 6x\bar{x} \bigg[1+\sum_{n=1}^{\mathcal{N}=2} a_{n}^{\sigma}(\mu) C_{n}^{3/2}(\xi) \bigg].
\end{align}
By choosing the above twist-3 distribution amplitudes as input parameters, we can obtain the complete TFFs.

\section{Numerical Analysis And Discussions}\label {sec:III}
For numerical calculations, we provide the following input parameters: the meson masses $m_{D^0}=1864.84\pm0.05~\rm{MeV}$, $m_{D^-}=1869.66\pm0.05~\rm{MeV}$, and $m_{a_0(1450)}=1439\pm34~\rm{MeV}$, the quark masses $m_c(\bar m_c)=1273\pm20~\rm{MeV}$, $m_d=4.67^{+0.48}_{-0.17}~\rm{MeV}$ and $m_u=2.16\pm0.07~\rm{MeV}$ at $\mu=2~\rm{GeV}$. The decay constant $f_D=208.4\pm1.5~\rm{MeV}$~\cite{ParticleDataGroup:2016lqr} and $\bar{f}_{a_0(1450)}=460\pm50~\rm{MeV}$ at $\mu_0 = 1\ \rm{GeV}$~\cite{Cheng:2005nb}. For the $D\to a_0(1450)$ transition, we also provide corresponding energy scale: $\mu_{k} = \sqrt{m_D^2 - m_c^2} \simeq 1.4\ \text{GeV}$.
\begin{figure}[t]
\begin{center}
\includegraphics[width=0.6\textwidth]{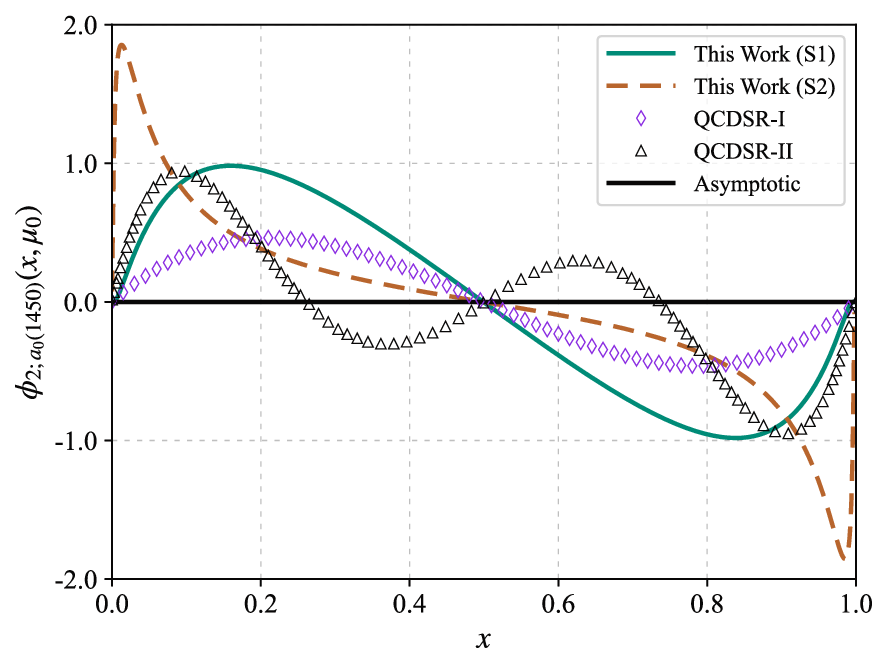}
\end{center}
\caption{Behavior of twist-2 LCDAs for $a_0(1450)$ at the $\mu_0 = 1~\rm{GeV}$. As a comparison, the QCDSR-I and  QCDSR-II (I and II in QCDSR~\cite{Cheng:2005nb} correspond to the orders of Gegenbauer moments $\mathcal{N}=1$ and $\mathcal{N}=3$, respectively) are also presented.}
\label{Fig:DA}
\end{figure}
As discussed in Section~\ref{sec:II} regarding the twist-2 LCDA model parameters, we adopt $a_1(\mu_0) = -0.58\pm0.12$ and $\left\langle \textbf{k}_{\perp}^2\right\rangle^{1/2} =0.37~\rm{GeV}^2$ for the first scheme. For the second scheme, an additional condition $a_3(\mu_0) = -0.49\pm0.15$ is imposed~\cite{Cheng:2005nb}. Based on these model parameter choices, the LCDAs of $a_{0}(1450)$ are determined. At the same time, we need to use renormalization group evolution (RGE) to evolve these parameters to corresponding energy scale $\mu_k$~\cite{Ball:2006nr,Ball:2004ye}.

At the initial energy scale $\mu_0=1~\rm{GeV}$, the model parameters of two LCDA schemes for light scalar state $a_0(1450)$ are presented, based on the typical values of $a_{1}(\mu_0)$ and $\left\langle \textbf{k}_{\perp}^2\right\rangle^{1/2}$,
\begin{align}
& A^{\rm (S1)}=390.35, && \beta^{\rm (S1)}=-0.724; \nonumber \\
& A^{\rm (S2)}=-20.346, && \beta^{\rm (S2)}=1.240, && \alpha^{\rm (S2)} =-0.878.
\end{align}
The corresponding LCDA behavior is presented in Fig.~\ref{Fig:DA}. It can be seen that our predictions for twist-2 LCDA within two schemes all exhibit an antisymmetric behavior, which along with the results from QCDSRs. Our result of S1 is also identical in behavior to QCDSR-I~\cite{Cheng:2005nb}. However, there are some differences, which may be due to the different models used: we construct twist-2 LCDA through the LCHO model, while QCDSRs adopts the Gegenbauer moment polynomial expansion. Furthermore, in the calculation of subsequent observations, we find that the numerical results of QCDSR-I and QCDSR-II are almost identical. Thus in the following discussion, we will only adopt QCDSR-II for analysis.

Using the definition Eq.~(\ref{moment}), we calculated moments $\langle\xi^{n}\rangle|_{\mu_0}^{\mathrm{(S1,S2)}}$ with $n=(1,3,5)$ and corresponding Gegenbauer moments $a_n^{(\rm{S1,S2})}(\mu_0)$. Since the zeroth-order Gegenbauer moment of $a_0(1450)$ is zero, and even-order coefficients are highly suppressed, under the approximation of equal constituent quark masses $(m_1 =m_2)$, even-order coefficients are equal to zero. Thus, the twist-2 LCDA of $a_0(1450)$ is dominated by odd-order Gegenbauer moments,
\begin{align}
&\langle\xi^{1} \rangle|^{\rm (S1)}_{\mu_0}=-0.348^{+0.072}_{-0.072},
~~~\langle\xi^{1} \rangle|^{\rm (S2)}_{\mu_0}=-0.348^{+0.072}_{-0.072},
\nonumber \\
&\langle\xi^{3} \rangle|^{\rm (S1)}_{\mu_0}=-0.160^{+0.033}_{-0.033},
~~~\langle\xi^{3} \rangle|^{\rm (S2)}_{\mu_0}=-0.242^{+0.059}_{-0.060},
\nonumber \\
&\langle\xi^{5} \rangle|^{\rm (S1)}_{\mu_0}=-0.092^{+0.019}_{-0.019},
~~~\langle\xi^{5} \rangle|^{\rm (S2)}_{\mu_0}=-0.188^{+0.051}_{-0.052},
\nonumber \\
&a_{1}^{\rm (S1)}(\mu_0)=-0.580^{+0.120}_{-0.120},
~~~\hspace{0.2em}a_{1}^{\rm (S2)}(\mu_0)=-0.580^{+0.120}_{-0.120},
\nonumber \\
&a_{3}^{\rm (S1)}(\mu_0)=-0.046^{+0.011}_{-0.011},
~~~\hspace{0.2em}a_{3}^{\rm (S2)}(\mu_0)=-0.488^{+0.148}_{-0.153},
\nonumber \\
&a_{5}^{\rm (S1)}(\mu_0)=+0.013^{+0.002}_{-0.002},
~~~\hspace{0.2em}a_{5}^{\rm (S2)}(\mu_0)=-0.371^{+0.148}_{-0.131}.
\end{align}
Meanwhile, we also present the $\langle\xi^{n} \rangle|^{\mathrm{(S1,S2)}}_{\mu_k}$ moments and Gegenbauer moments $a_n^{(\rm{S1,S2})}(\mu_k)$ at the corresponding energy scale $\mu_{k}=1.4~\rm{GeV}$,
\begin{align}
&\langle\xi^{1} \rangle|^{\rm (S1)}_{\mu_k}=-0.304^{+0.065}_{-0.058},
~~~\langle\xi^{1} \rangle|^{\rm (S2)}_{\mu_k}=-0.249^{+0.065}_{-0.065},
\nonumber \\
&\langle\xi^{3} \rangle|^{\rm (S1)}_{\mu_k}=-0.137^{+0.029}_{-0.026},
~~~\langle\xi^{3} \rangle|^{\rm (S2)}_{\mu_k}=-0.173^{+0.044}_{-0.045},
\nonumber \\
&\langle\xi^{5} \rangle|^{\rm (S1)}_{\mu_k}=-0.078^{+0.017}_{-0.014},
~~~\langle\xi^{5} \rangle|^{\rm (S2)}_{\mu_k}=-0.135^{+0.034}_{-0.034},
\nonumber \\
&a_{1}^{\rm (S1)}(\mu_k)=-0.507^{+0.109}_{-0.096},
~~~\hspace{0.2em}a_{1}^{\rm (S2)}(\mu_k)=-0.415^{+0.108}_{-0.108},
\nonumber \\
&a_{3}^{\rm (S1)}(\mu_k)=-0.035^{+0.006}_{-0.006},
~~~\hspace{0.2em}a_{3}^{\rm (S2)}(\mu_k)=-0.348^{+0.085}_{-0.090},
\nonumber \\
&a_{5}^{\rm (S1)}(\mu_k)=+0.009^{+0.008}_{-0.015},
~~~\hspace{0.2em}a_{5}^{\rm (S2)}(\mu_k)=-0.277^{+0.070}_{-0.052}.
\end{align}
For orders $n=(1,3,5)$, the moments obtained from two LCDA schemes generally exhibit a linear relationship, where the absolute values decrease as $n$ increases, and the absolute values of moments from first scheme are larger. Specifically, at order $n=1$, the $\langle\xi^{1}\rangle|^{\mathrm{(S1/S2)}}_{\mu_0}$ values from two LCDA schemes are numerically the same, whereas for the $n=(3,5)$ orders, differences between moments from two LCDA schemes increase significantly with the order, indicating that factor $(2x-1)^n$ of higher-order moments is more sensitive to the value of $x$.
\begin{table}
    \caption{Numerical results of TFFs for the $D\to a_0(1450)$ transition at the large recoil point {\it i.e.} $f_+^{\rm (S1,S2)}(0)$ and $f_{-}(0)$. To make a comparison, we also listed  QCDSR-II~\cite{Cheng:2005nb}, CLFQM~\cite{Verma:2011yw}, RQM~\cite{Galkin:2025emi} and LCSR~\cite{Huang:2021owr} predictions.}
    \label{The TFFs value}
    \centering
    \renewcommand{\arraystretch}{1}
    \begin{tabular}{l l l}
        \hline
         ~~~~~~& $f_+(0)$ ~~~~~~&$~~f_{-}(0)$ \\
        \hline
        This work ($\mathrm{S1}$) ~~~~~~~& $0.836_{-0.119}^{+0.116}$ ~~~~~~~&$~~0.630_{-0.077}^{+0.078}$ \\
        This work ($\mathrm{S2}$) ~~~~~~~& $0.767_{-0.105}^{+0.106}$ ~~~~~~~&$~~0.630_{-0.077}^{+0.078}$ \\
        QCDSR-II~\cite{Cheng:2005nb} ~~~~~~~& $0.691$ ~~~~~~~& $~~--$ \\
        CLFQM~\cite{Verma:2011yw} ~~~~~~~& $0.51_{+0.01-0.02}^{-0.01+0.01}$ ~~~~~~~& $~~--$ \\
        RQM~\cite{Galkin:2025emi} ~~~~~~~& $0.719$ ~~~~~~~& $-1.391$ \\
        LCSR~\cite{Huang:2021owr} ~~~~~~~& $0.94_{-0.03}^{+0.02}$ ~~~~~~~& $-0.94_{-0.03}^{+0.02}$ \\
        \hline
    \end{tabular}
\end{table}
In addition, the LCDA $\phi_{3;a_0}^{p,\sigma}(x,\mu)$ of $a_0(1450)$ twist-3 distribution amplitude, we adopt Gegenbauer moments at the energy scale $\mu_{k}=1.4~\rm{GeV}$ as follows~\cite{Han:2013zg}:
\begin{align}
&a_{2}^{p}(\mu_{k})=0.236 \pm{0.008}, &a_{4}^{p}(\mu_{k})=0.504 \pm{0.159}, \nonumber \\
&a_{2}^{\sigma}(\mu_{k})=0.009 \pm{0.001}, &a_{4}^{\sigma}(\mu_{k})=0.043 \pm{0.008},
\end{align}

With the resulatant $a_0(1450)$-state twist-2 and twist-3 LCDAs, we can further to calculate the $D\to a_0(1450)$ TFFs. To determine the continuum threshold and Borel windows, we always take the two criteria: (1) the contributions from the continuum and higher excited states are less than 30\%; (2) the dependence of form factors on the Borel parameter is weak. So we can determine the values of $s_0$ and $M^2$ to be $2.9 \pm0.05 {~\rm GeV}^2$ and $15 \pm0.05 {~\rm GeV}^2$ for $f_+^{\rm{(S1,S2)}}(q^2)$, and $3 \pm0.05 {~\rm GeV}^2$ and $5 \pm0.05 {~\rm GeV}^2$ for $f_- (q^2)$, respectively. Then, we obtain the $D \to a_0(1450)$ transition process TFFs numerical results in the large recoil region, which are presented in Table~\ref{The TFFs value}. At the same time, we also present the predictions of QCDSR-II~\cite{Cheng:2005nb}, CLFQM~\cite{Verma:2011yw}, RQM~\cite{Galkin:2025emi} and LCSR~\cite{Huang:2021owr} for comparison. It is found that our numerical results $f_+^{ (\mathrm{S2})}(0)$ are relatively closest to RQM. The prediction results of QCDSR-II~\cite{Cheng:2005nb} are also within our uncertainty range.

\begin{table}
    \caption{Under two LCDA schemes, the masses of $D$-meson and their resonances, fitting parameters $a_i$ with $i=(1,2)$, and the goodness of fit $\Delta$ corresponding to form factors $f_{\pm}^{\rm{(S1,S2)}} (q^2)$ are provided, with all input parameters set to their central values.}
    \label{The fit parameters}
    \centering
    \renewcommand{\arraystretch}{1}
    \begin{tabular}{l l l}
        \hline
         ~~~~~~~~~~~~~~&$~~f_+ (q^2)$ ~~~~~~~~~~ &$~~f_{-} (q^2)$ \\
        \hline
        $m_{R,i}$ ~~~~~~~~~~ &$~~2.00685$ ~~~~~~~~~~ &~~$2.343$ \\
        $a_1^{(\mathrm{S1})}$ ~~~~~~~~~~ &$-7.418$ ~~~~~~~~~~ &$-12.430$ \\
        $a_2^{(\mathrm{S1})}$ ~~~~~~~~~~ &$~~145.839$ ~~~~~~~~~~ &~~$275.236$ \\
        $\Delta^{(\mathrm{S1})}$ ~~~~~~~~~~ &$~~0.117\times 10^{-3}$ ~~~~~~~~~~ &~~$0.219\times 10^{-3} $ \\
        \hline
        $m_{R,i}$ ~~~~~~~~~~ &~~$2.00685$ ~~~~~~~~~~ &~~$2.343$ \\
        $a_1^{(\mathrm{S2})}$ ~~~~~~~~~~ &$-5.077$ ~~~~~~~~~~ &$-12.430$ \\
        $a_2^{(\mathrm{S2})}$ ~~~~~~~~~~ &~~$146.277$ ~~~~~~~~~~ &~~$275.236$ \\
        $\Delta^{(\mathrm{S2})}$ ~~~~~~~~~~ &~~$0.132\times 10^{-3} $ ~~~~~~~~~~ &~~$0.219\times 10^{-3} $ \\
        \hline
    \end{tabular}
\end{table}

Our result for $f_-(0)$ differs from those given from QCDSR-II~\cite{Cheng:2005nb} and CLFQM~\cite{Verma:2011yw}. This discrepancy may be attributed to differences in the methods and parameters. In particular, the substantial discrepancy with LCSR~\cite{Huang:2021owr} prediction arises from differences in our choice of twist-2 LCDA model and formulation of TFFs. Currently, there are some variations among theoretical groups in predicting the numerical results of $f_- (0)$ at the large-recoil point. Moreover, we can further investigate the overall behavioral patterns of TFFs to evaluate the rationality of numerical results.

\begin{figure}[t]
\begin{center}
\includegraphics[width=0.5\textwidth]{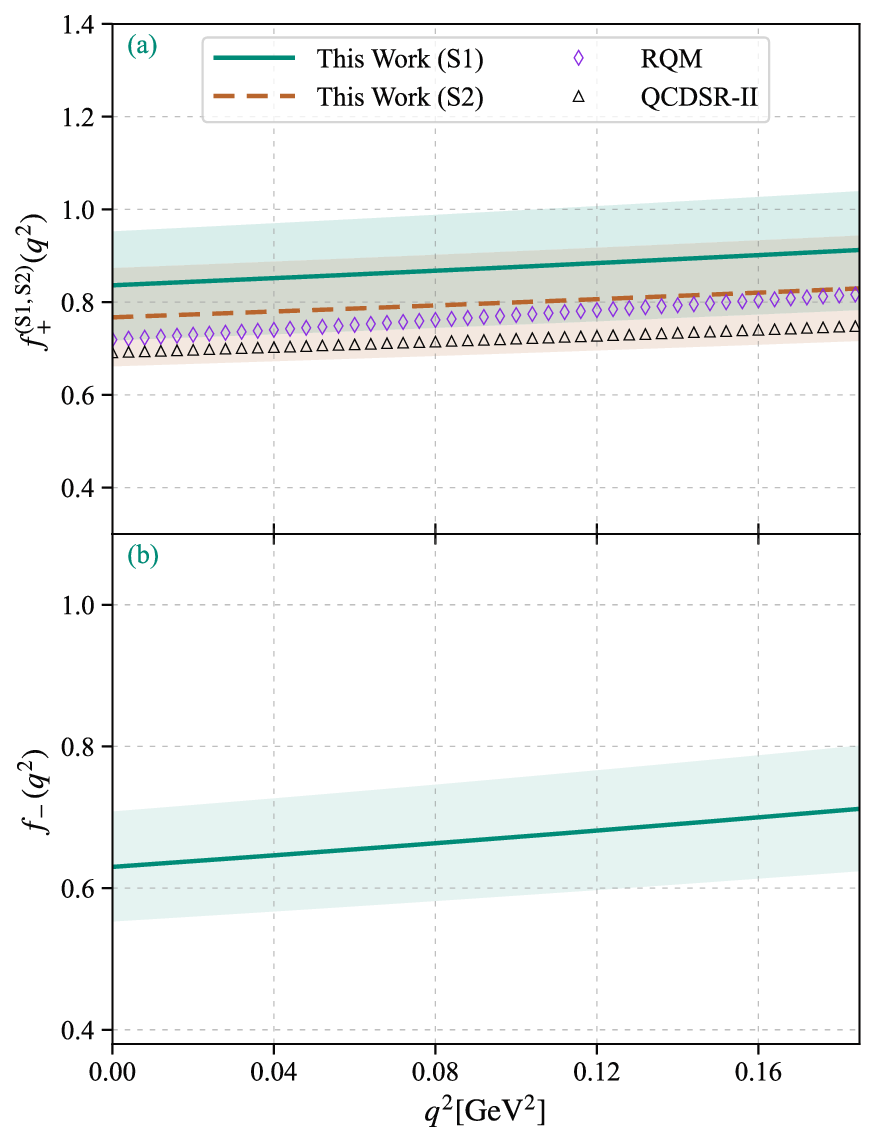}
\end{center}
\caption{Behaviors of $D \to a_0(1450)$ TFFs $f_{\pm} (q^2)$ in the entire $q^2$-region, where solid lines represent the central values and shaded regions represent the uncertainty ranges. For comparison, predictions from RQM~\cite{Galkin:2025emi} and QCDSR-II~\cite{Cheng:2005nb} are also provided.}
\label{Fig:TFFs}
\end{figure}
\begin{figure}[t]
\begin{center}
\includegraphics[width=0.98\textwidth]{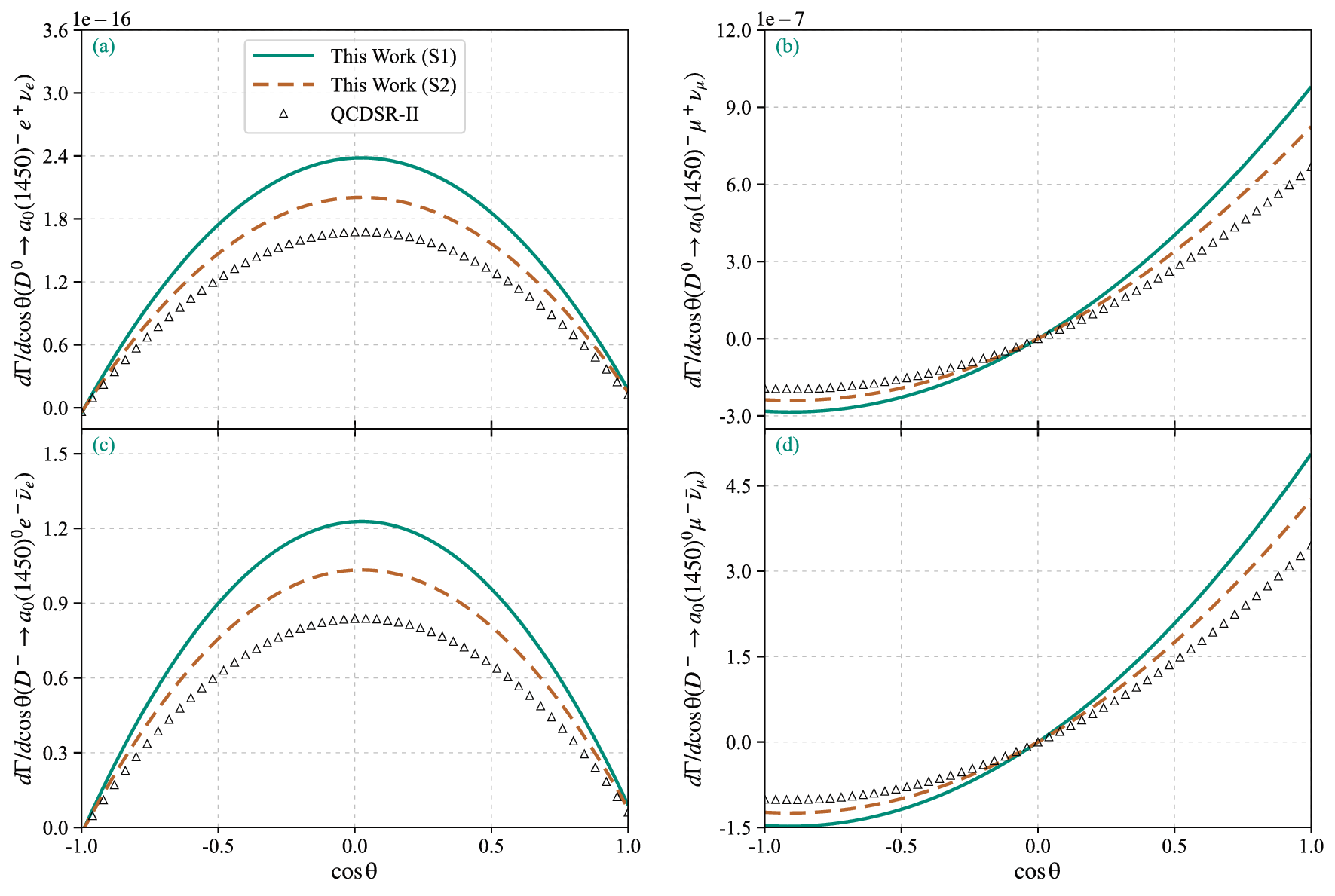}
\caption{Angular distribution of ${d\Gamma}/{d\cos\theta}$ with respect to $\cos\theta$ for the decay channels $D^0 \to a_0(1450)^- \ell^+ \nu_\ell $ and $D^- \to a_0(1450)^0 \ell^- \bar{\nu}_\ell $ with $\ell=(e,\mu$), where $q^2=10^{-5}~{\rm GeV}^2$. As make a comparison, the prediction results of QCDSR-II~\cite{Cheng:2005nb} is also given.}
\label{Fig: angular distribution}
\end{center}
\end{figure}

Since the LCSR approach is mainly applicable to low and intermediate $q^2$-regions, we need to adopt simplified series expansion (SSE) to fit the complex analytical results and then extrapolate them to the entire $q^2$-region to obtain a reasonable behavior of TFF for transition process $D \to a_0(1450)$ over the full physical range {\it i.e.} $0 \leq q^2 \leq ( m_D-m_{a_0(1450)})^2$. The SSE is a fast convergent series on $z(t)$-expansion, that its expansion can be written as follows~\cite{Fu:2018yin},
\begin{align}
f_i(q^2) = P_i(q^2) \sum_{k=0,1,2} a_k^i[z(q^2)-z(0)]^k,
\end{align}
where $f_i(q^2)$ stands for $D \to a_0(1450)$ TFFs, $a_k^i$ are the coefficients, and $P_i(q^2) = \big(1-{q^2}/{m_{R,i}^2} \big)^{-1}$, $z(t) = ({\sqrt{t_+-t}-\sqrt{t_+-t_0}})/({\sqrt{t_+-t}+\sqrt{t_+-t_0}})$, with $t_+=(m_D - m_{a_0(1450)})^2$, $t_{0}=t_+(1-\sqrt{1-(t_{-})/(t_+)})$. $P_i(q^2)$ is a simple pole corresponding to the first-order resonance in the spectrum and can be used to account for the low-lying resonance, $m_{R,i}$ is the resonance of $D$-meson. The masses of low-lying $D$-meson resonances are mainly determined by their $J^P$ quantum number states, which are $m_{D^*}(2007)^0$ and $m_{D^*_0}(2300)$ here. Free parameters $a_{1}^i$ and $a_{2}^i$ are also determined to make the goodness of fit $\Delta$ of TFF as small as possible. Conventionally, we require $\Delta < 1\%$, and $\Delta$ is used to measure the extrapolation quality. Meanwhile, we also present the fitting parameters for TFFs in Table~\ref{The fit parameters}.

After extrapolating $D \to a_0(1450)$ TFFs to the whole physical $q^2$-region, the behavior of $f_+^{ (\rm{S1,S2})}(q^2)$ and $f_{-} (q^2)$ are present in Fig.~\ref{Fig:TFFs}, where the $f_+^{ (\rm{S1,S2})}(q^2)$ is correspond to the two types of $a_0(1450)$-state twist-2 LCDA. The behavior of $f_+^{ (\mathrm{S1,S2})}(q^2)$ exhibits a slowly increasing trend within the uncertainty range across entire $q^2$-region. This is in good agreement with the predictions from the QCDSR-II~\cite{Cheng:2005nb} theory group and RQM~\cite{Galkin:2025emi} collaboration, indicating that our results are reasonable. For $f_{-} (q^2)$, since its expression does not include contributions from twist-2 LCDA, the different twist-2 LCDA schemes have no impact on $f_{-} (q^2)$.
\begin{figure*}
\begin{center}
\includegraphics[width=0.98\textwidth]{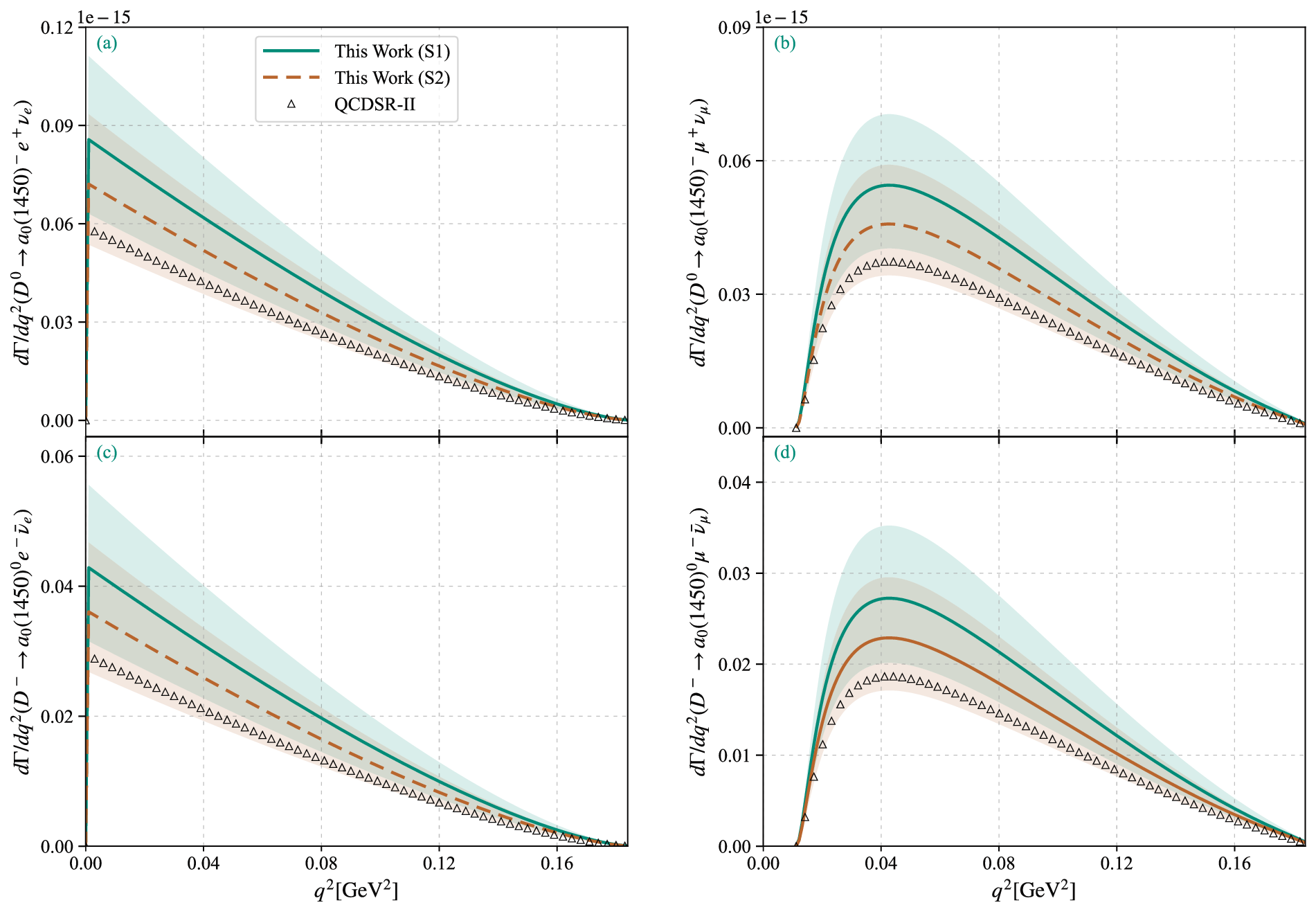}
\caption{Decay widths for the semileptonic decay channels $D^0 \to a_0(1450)^- \ell^+ \nu_\ell $ and $D^- \to a_0(1450)^0 \ell^- \bar{\nu}_\ell $ with $\ell=(e ,\mu)$. QCDSR-II~\cite{Cheng:2005nb} predictions are present for comparison. Shaded bands indicate uncertainties.}
\label{Fig: decay width}
\end{center}
\end{figure*}

Furthermore, the distribution of differential decay width with respect to angle $\cos\theta_\ell$, {\it i.e.} ${d\Gamma}/{d\cos\theta_\ell }$ at the point $q^2=10^{-5}~{\rm GeV}^2$ in the region $\cos\theta_\ell  \in[-1,1]$ are shown in Fig.~\ref{Fig: angular distribution}. The results show that ${d\Gamma}/{d\cos\theta_\ell }$ exhibits a quadratic dependence on $\cos\theta_\ell $, with coefficients primarily determined by the TFFs $f_+^{\rm{(S1,S2)}}(q^2)$, $f_- (q^2)$ and lepton mass $m_\ell $. For decay channel $D^0\to a_0(1450)^- \mu^+\nu_{\mu}$ and $D^-\to a_0(1450)^0 \mu^-\bar{\nu}_{\mu}$, the angular distribution increases monotonically with $\cos\theta_\ell $ and shows a nonzero intercept at the endpoints $\cos\theta_\ell = \pm 1$. This occurs because the angular distribution contains a term linear in $\cos\theta_\ell $, which is proportional to ${m_\ell ^2}/{q^2}$, leading to a finite value of ${d\Gamma}/{d\cos\theta_\ell }$ at the endpoints when $q^2=10^{-5}~{\rm GeV}^2$. In contrast, for decay channel $D^0\to a_0(1450)^- e^+\nu_{e}$ and $D^-\to a_0(1450)^0 e^- \bar{\nu}_{e}$, the extremely small electron mass ($m_e \approx 0.5 \rm~{MeV}$) makes angular distribution over $\cos\theta_\ell  \in[-1,1]$ nearly a symmetric parabola. These results demonstrate that the lepton mass has a important impact on the decay angular distribution in the low $q^2$-region.

As a further step, the decay widths and branching fractions for semileptonic $D \to a_0(1450) \ell \nu{_\ell}$ decay can be calculated. By taking the CKM matrix element $|V_{cd}|=0.221 \pm{0.008}$ from the PDG~\cite{ParticleDataGroup:2020ssz}, we obtain the differential decay widths for $D \to a_0(1450) \ell \nu{_\ell}$ with $\ell=(e,\mu)$. Their behaviors are shown in Fig.~\ref{Fig: decay width}, together with the QCDSR-II prediction~\cite{Cheng:2005nb} for comparison. Since the kinematic $q^2$-region of $D \to a_0(1450) \ell \nu_\ell $ decay is much smaller than that of $B \to a_0(1450) \ell \nu{_\ell}$ decay, we separately consider the differential decay widths for electron and muon channels. It can be seen from Fig.~\ref{Fig: decay width} that the behavioral trends of our predicted differential decay widths for electrons and muons under two LCDA schemes are basically consistent, though there exist slight differences in numerical results. Compared with the first scheme, the second scheme gives smaller numerical results. This is because the TFF $f^{  (\rm{S2})}_+(q^2)$ obtained in second scheme is slightly lower than $f^{ (\rm{S1})}_+(q^2)$ from the first scheme. Therefore, under the same threshold parameter $s_0$ and Borel window $M^2$, the differential decay widths for the electron and muon channels in second scheme $f^{(\rm{S2})}_+(q^2)$ are also smaller.

After taking the lifetimes of $D^0$ and $D^-$ mesons from the PDG, namely $\tau_{D^0}=(0.410 \pm{0.001})$ ps and $\tau_{D^-}=(1.033\pm{0.005})$ ps, we also calculated the branching fractions of $D^0 \to a_0(1450)^{-} \ell^{+} \nu_\ell $ and $D^- \to a_0(1450)^{0} \ell^{-} \bar{\nu}_\ell $ with $(\ell= e,\mu)$. The specific calculation results are listed in Table~\ref{table:Br}. It can be observed that branching fractions for all decay channels are of the order $10^{-6}$, consistent with results reported by theoretical groups using LCSR~\cite{Huang:2021owr} and CLFQM~\cite{Cheng:2017pcq}.

\begin{table}[t]
\begin{center}
\renewcommand{\arraystretch}{1.1}
\footnotesize
\caption{Branching fractions (in unit: $10^{-6}$) for the decay channels of $D^0 \to a_0(1450)^- \ell^+ \nu_\ell $ and $D^- \to a_0(1450)^0 \ell^- \bar{\nu}_\ell $ with $\ell=(e,\mu)$ within uncertainties under two twist-2 LCDA schemes are presented, with the predictions from LCSR~\cite{Huang:2021owr}, QCDSR-II~\cite{Cheng:2005nb} and CLFQM~\cite{Cheng:2017pcq} provided for comparison.}
\label{table:Br}
\begin{tabular}{l l l l l}
\hline
~~~~~~~~~~~~~~~~~~& $D^0\to a_0(1450)^- e^+\nu_{e}$ ~~~~& $D^0\to a_0(1450)^- \mu^+\nu_{\mu} $ ~~~~& $D^-\to a_0(1450)^0 e^-\bar{\nu}_{e} $ ~~~~& $D^-\to a_0(1450)^0 \mu^-\bar{\nu}_{\mu} $ \\
\hline
This work $(\mathrm{S1})$          ~~~&$4.16^{+1.23}_{-1.10}$            ~~~&$1.91^{+2.27}_{-0.47}$       ~~~&$5.24^{+1.55}_{-1.38}$       ~~~&$2.41^{+2.86}_{-0.59}$\\
This work $(\mathrm{S2})$   ~~~&$3.48^{+1.03}_{-0.89}$            ~~~&$1.91^{+1.27}_{-0.26}$   ~~~&$4.39^{+1.29}_{-1.12}$       ~~~&$2.41^{+2.01}_{-0.13}$\\
QCDSR-II~\cite{Cheng:2005nb}         ~~~&$2.83$            ~~~&$1.92$       ~~~&$3.56$            ~~~&$2.43$\\
LCSR~\cite{Huang:2021owr}         ~~~&$3.14$            ~~~&$2.01$       ~~~&$4.28$            ~~~&$2.76$\\
CLFQM~\cite{Cheng:2017pcq}         ~~~&$--$            ~~~&$--$       ~~~&$5.4^{+0.05}_{-0.05}$            ~~~&$3.8^{+0.03}_{-0.03}$\\
\hline
\end{tabular}
\end{center}
\end{table}

For decay channels $D^- \to a_0(1450)^{0} e^{-}\bar{\nu}_e$ and $D^0 \to a_0(1450)^{-}e^{+} \nu_e$, the numerical results calculated via two LCDA schemes are closer to LCSR, while exhibiting discrepancies from the CLFQM results for $D^- \to a_0(1450)^{0} e^{-}\bar{\nu}_e$ and $D^- \to a_0(1450)^{0} \mu^{-}\bar{\nu}_\mu$. Notably, the central values of branching fractions for $D^0 \to a_0(1450)^{-}\mu^{+} \nu_{\mu}$ and $D^- \to a_0(1450)^{0} \mu^{-}\bar{\nu}_{\mu}$ derived from two LCDA schemes and QCDSR-II~\cite{Cheng:2005nb} are identical, differing only marginally in their theoretical uncertainties. Furthermore, the central values from both LCSR and CLFQM calculations are within our allowed uncertainty range.

\begin{table}[t]
\renewcommand{\arraystretch}{1.1}
\footnotesize
\caption{Integrated results for three angular observables of the semileptonic decay $D^0 \to a_0(1450)^- \ell^+ \nu_\ell $ with $\ell=(e,\mu)$, evaluated for electrons and muons under two twist-2 LCDA schemes, are presented.}
\label{table:Observables}
\begin{tabular}{l c c}
\hline
 ~~~~~~& This work~$(\mathrm{S1})$ ~~~~~~& This work~$(\mathrm{S2})$ \\
\hline
${\cal A}_{\rm FB}^{D^0 \to a_0(1450)^- e^+ \nu_{e}}(10^{-6})$   ~~~~~~& $5.46^{+2.54}_{-1.90}$ ~~~~~~& $5.47^{+2.47}_{-1.89}$  \\
${\cal A}_{\rm FB}^{D^0 \to a_0(1450)^-\mu^+ \nu_{\mu}}(10^{-2})$  ~~~~~~& $3.66^{+1.67}_{-1.26}$ ~~~~~~& $3.66^{+1.62}_{-1.23}$  \\
${\cal A}_{\lambda \ell}^{D^0 \to a_0(1450)^- e^+ \nu_{e}}(10^{-1})$ ~~~~~~& $0.18^{+0.31}_{-0.31}$ ~~~~~~& $0.18^{+0.36}_{-0.36}$  \\
${\cal A}_{\lambda \ell}^{D^0 \to a_0(1450)^-\mu^+ \nu_{\mu}}(10^{-2})$ ~~~~~~& $0.02^{+0.05}_{-0.06}$ ~~~~~~& $0.02^{+0.05}_{-0.06}$  \\
${\cal F}_{\rm H}^{D^0 \to a_0(1450)^- e^+ \nu_{e}}(10^{-5})$ ~~~~~~& $1.60^{+0.73}_{-0.54}$ ~~~~~~& $1.60^{+0.71}_{-0.54}$  \\
${\cal F}_{\rm H}^{D^0 \to a_0(1450)^-\mu^+ \nu_{\mu}}(10^{-2})$ ~~~~~~& $9.03^{+4.07}_{-3.06}$ ~~~~~~& $9.06^{+3.95}_{-3.00}$  \\
\hline
\end{tabular}
\end{table}
Finally, we calculated three angular observables of the semileptonic decay $D\to a_0(1450) \ell \nu{_\ell}$: the forward-backward asymmetry ${\cal A}_{\rm FB}(q^2)$, lepton polarization asymmetry ${\cal A}_{\lambda_\ell}$ and $q^2$-differential flat term ${\cal F}_{\rm H}(q^2)$. Since the integrated results for $D^0 \to a_0(1450)^- \ell^+ \nu{_\ell}$ and $D^- \to a_0(1450)^0 \ell^- \bar{\nu}_\ell $ decay channels are almost identical, we only present the results for $D^0 \to a_0(1450)^- \ell^+ \nu{_\ell}$ here. As shown in Table~\ref{table:Observables}, three angular observables calculated using two twist-2 LCDA schemes show very similar results, with only slight differences. The integrated ${\cal A}_{\rm FB}$ for electron is very small ($m_{e} \approx 0.5 \rm~{MeV}$), which may be because in the weak interaction, the helicity of lepton is highly dependent on the mass, and the mass of electron is almost 0. The mass of muon ($m_{\mu} \approx 105.7~\rm{MeV}$) is relatively large, so there is a slight difference in the numerical value of ${\cal A}_{\rm FB}(q^2)$ between the two LCDA schemes. In addition, distinct results between the ${\cal A}_{\lambda_\ell}(q^2)$ and ${\cal F}_{\rm H}(q^2)$ are observed. Overall, the ${\cal A}_{\rm FB}(q^2)$ and ${\cal F}_{\rm H}(q^2)$ are proportional to square of the lepton mass, while ${\cal A}_{\lambda_\ell}(q^2)$ is inversely proportional to lepton mass.

\section{Summary}\label {Sec:IV}
Motivated by the possibility that the scalar state near $1.5~\rm{GeV}$ may have a diquark state, we assume that the $a_0(1450)$ is a ground $q\bar{q}$ state in this work. We attempt to discuss the physical observables from the semileptonic decay process $D \to a_0(1450)\ell \nu_\ell $ with $\ell=(e,\mu)$, in order to further help us judge the rationality of the diquark state perspective.

For the TFFs of $D \to a_0(1450)$, we employ the LCSR method for calculation. As the most important nonperturbative input, the twist-2 LCDA is described by constructing two different forms of LCHO model, which provides a phenomenological perspective on the momentum fraction distribution of the partons inside the $a_0(1450)$. The specific behavior is shown in Fig.~\ref{Fig:DA}. Furthermore, we obtain the $\langle\xi^{n} \rangle |_{\mu}$ and Gegenbauer moments $a_n(\mu)$ at scale $\mu_0=1~{\rm GeV}$ and $\mu_k=1.4~{\rm GeV}$. Subsequently, by incorporating these two twist-2 LCDAs, we calculate the TFF $f_+(q^2)$ at large recoil region $q^2=0$ and its behavior in the low and intermediate $q^2$ region. The results show good consistency with the predictions from the QCDSR-II~\cite{Cheng:2005nb}, RQM~\cite{Galkin:2025emi} and LCSR~\cite{Huang:2021owr} within the allowed uncertainties. Due to the differences in parameters and methods, there are certain variations in TFF $f_-(q^2)$. Then, by using the global behavior of these two TFFs, we predicted angular distribution of the differential decay width ${d\Gamma}/{d\cos\theta_\ell}$ in the range $\cos\theta_\ell \in [-1,1]$, as shown in Fig.~\ref{Fig: angular distribution}. The differential
decay widths and branching fractions of $D \to a_0(1450)\ell \nu_\ell $ are also obtained, which are presented in Fig.~\ref{Fig: decay width} and Table~\ref{table:Br}, respectively. Our results are consistent in order of magnitude with those from QCDSR-II~\cite{Cheng:2005nb}, LCSR~\cite{Huang:2021owr} and CLFQM~\cite{Cheng:2017pcq}. Finally, we calculate the integrated results of three angular observables: the forward-backward asymmetry ${\cal A}_{\rm FB}$, lepton polarization asymmetry ${\cal A}_{\lambda_\ell}$ and $q^2$-differential flat term ${\cal F}_{\rm H}(q^2)$, which are listed in Table~\ref{table:Observables}.

Overall, under the assumption of a diquark state, our predictions for the physical observables of the semileptonic decay $D \to a_0(1450)\ell \nu_\ell $ show good agreement with existing theoretical studies. The internal structure of the scalar family remains a highly debated topic. We hope that our results can provide a useful reference for future experimental searches for the semileptonic decay $D \to a_0(1450)\ell \nu_\ell $, and contribute to a deeper understanding of the internal structure of the light scalar state $ a_0(1450)$.

\section{Acknowledgments}
This work was supported in part by the National Natural Science Foundation of China under Grant No.12265010, the Project of Guizhou Provincial Department of Science and Technology under Grant No.MS[2025]219, No.CXTD[2025]030.


\begin{thebibliography}{99}

\bibitem{Cheng:2005nb}
H.~Y.~Cheng, C.~K.~Chua and K.~C.~Yang [ADS Abstract Service]
Charmless hadronic $B$ decays involving scalar mesons: Implications to the nature of light scalar mesons,
\href{https://doi.org/10.1103/PhysRevD.73.014017}
{Phys. Rev. D \textbf{73} (2006) 014017}.
[\href{https://arxiv.org/pdf/hep-ph/0508104}
{hep-ph/0508104}]



\bibitem{Rui:2018mxc}
Z.~Rui, Y.~Q.~Li and J.~Zhang,
Isovector scalar $a_0(980)$ and $a_0(1450)$ resonances in the $B \to \psi (K\bar{K},\pi\eta)$ decays,
\href{https://doi.org/10.1103/PhysRevD.99.093007}
{Phys. Rev. D \textbf{99} (2019) 093007}.
[\href{https://arxiv.org/pdf/1811.12738}
{arXiv:1811.12738}]


\bibitem{Chai:2021pyp}
J.~Chai, S.~Cheng and A.~J.~Ma,
Probing isovector scalar mesons in the charmless three-body $B$ decays,
\href{https://doi.org/10.1103/PhysRevD.105.033003}
{Phys. Rev. D \textbf{105} (2022) 033003}.
[\href{https://arxiv.org/pdf/2109.00664}
{arXiv:2109.00664}]

\bibitem{Guo:2022xqu}
D.~Guo, W.~Chen, H.~X.~Chen, X.~Liu and S.~L.~Zhu,
Newly observed $a_0(1817)$ as the scaling point of constructing the scalar meson spectroscopy,
\href{https://doi.org/10.1103/PhysRevD.105.114014}
{Phys. Rev. D \textbf{105} (2022) 114014}.
[\href{https://arxiv.org/pdf/2204.13092}
{arXiv:2204.13092}]

\bibitem{Han:2013zg}
H.~Y.~Han, X.~G.~Wu, H.~B.~Fu, Q.~L.~Zhang and T.~Zhong,
Twist-3 Distribution Amplitudes of Scalar Mesons within the QCD Sum Rules and Its Application to the $B \to S$ Transition Form Factors,
\href{https://doi.org/10.1140/epja/i2013-13078-7}
{Eur. Phys. J. A \textbf{49} (2013) 78}.
[\href{https://arxiv.org/pdf/1301.3978}
{arXiv:1301.3978}]


\bibitem{Jaffe:1976ig}
R.~L.~Jaffe,
Multi-Quark Hadrons. 1. The Phenomenology of $Q^2 \bar{Q}^2$ Mesons,
\href{https://doi:10.1103/PhysRevD.15.267}
{Phys. Rev. D \textbf{15} (1977) 267}.


\bibitem{Weinstein:1983gd}
J.~D.~Weinstein and N.~Isgur,
The $qq\bar{q} \bar{q}$ System in a Potential Model,
\href{https://doi:10.1103/PhysRevD.27.588}
{Phys. Rev. D \textbf{27} (1983) 588}.

\bibitem{Klempt:2021nuf}
E.~Klempt,
Scalar mesons and the fragmented glueball,
\href{https://doi.org/10.1016/j.physletb.2021.136512}
{Phys. Lett. B \textbf{820} (2021) 136512}.
[\href{https://arxiv.org/pdf/2104.09922}
{arXiv:2104.09922}]

\bibitem{Brito:2004tv}
T.~V.~Brito, F.~S.~Navarra, M.~Nielsen and M.~E.~Bracco,
QCD sum rule approach for the light scalar mesons as four-quark states,
\href{https://doi.org/10.1016/j.physletb.2005.01.008}
{Phys. Lett. B \textbf{608} (2005) 69-76}.
[\href{https://arxiv.org/abs/hep-ph/0411233}
{hep-ph/0411233}]


\bibitem{Klempt:2007cp}
E.~Klempt and A.~Zaitsev,
Glueballs, Hybrids, Multiquarks. Experimental facts versus QCD inspired concepts,
\href{https://doi.org/10.1016/j.physrep.2007.07.006}
{Phys. Rept. \textbf{454} (2007) 1-202}.
[\href{https://arxiv.org/abs/0708.4016}
{arXiv:0708.4016}]

\bibitem{BESIII:2018sjg}
M.~Ablikim \textit{et al}. [BESIII Collaboration],
Observation of the Semileptonic Decay $D^0 \to a_0(980)^- e^+ \nu_e$ and Evidence for $D^+ \to a_0(980)^0 e^+ \nu_e$,
\href{https://doi.org/10.1103/PhysRevLett.121.081802}
{Phys. Rev. Lett. \textbf{121} (2018) 081802}.
[\href{https://arxiv.org/abs/1803.02166}
{arXiv:1803.02166}]

\bibitem{BESIII:2021drk}
M.~Ablikim \textit{et al}. [BESIII Collaboration],
Study of light scalar mesons through $D^+_s \to \pi^0\pi^0e^+ \nu_e$ and $K^0_S K^0_S e^+ \nu_e$ decays,
\href{https://doi.org/10.1103/PhysRevD.105.L031101}
{Phys. Rev. D \textbf{105} (2022) L031101}.
[\href{https://arxiv.org/abs/2110.13994}
{arXiv:2110.13994}]

\bibitem{BESIII:2023wgr}
M.~Ablikim \textit{et al}. [BESIII Collaboration],
Study of the $f_0(980)$ and $f_0(500)$ Scalar Mesons through the Decay $D_s^+\to \pi^+\pi^-e^+\nu_e$,
\href{https://doi.org/10.1103/PhysRevLett.132.141901}
{Phys. Rev. Lett. \textbf{132} (2024) 141901}.
[\href{https://arxiv.org/abs/2303.12927}
{arXiv:2303.12927}]


\bibitem{BESIII:2023opt}
M.~Ablikim \textit{et al}. [BESIII Collaboration],
Studies of the decay $ D_s^+\to K^{+}K^{-}{\mu}^{+}{\nu}_{\mu } $,
\href{https://doi.org/10.1007/JHEP12(2023)072}
{JHEP \textbf{12} (2023) 072}.
[\href{https://arxiv.org/abs/2307.03024}
{arXiv:2307.03024}]

\bibitem{CLEO:2009ugx}
K.~M.~Ecklund \textit{et al}. [CLEO Collaboration],
Study of the semileptonic decay $D_s^+\to f_0(980) e^+ \nu$ and implications for $B_s^0\to J/\psi f_0$,
\href{https://doi.org/10.1103/PhysRevD.80.052009}
{Phys. Rev. D \textbf{80} (2009) 052009}.
[\href{https://arxiv.org/abs/0907.3201}
{arXiv:0907.3201}]

\bibitem{ParticleDataGroup:2020ssz}
P.~A.~Zyla $\textit{et al}.$ [Particle Data Group],
Review of Particle Physics,
\href{https://doi.org10.1093/ptep/ptaa104}
{PTEP \textbf{2020} (2020) 083C01}.

\bibitem{CrystalBarrel:1994arw}
Amsler, C. and others [Crystal Barrel Collaboration],
Observation of a new $I^G (J^{PC}) = 1^- (0^{++})$ resonance at $1450~\mathrm{MeV}$,
\href{https://doi.org/10.1016/0370-2693(94)91044-8}
{Phys. Lett. B \textbf{333} (1994) 277-282}.

\bibitem{Mathur:2006bs}
Mathur, Nilmani and Alexandru, A. and Chen, Y. and Dong, S. J. and Draper, Terrence and Horvath, I. and Lee, F. X. and Liu, K. F. and Tamhankar, S. and Zhang, J. B,
Scalar Mesons $a_0(1450)$ and $\sigma(600)$ from Lattice QCD,
\href{https://doi.org/10.1103/PhysRevD.76.114505}
{Phys. Rev. D. \textbf{76} (2007) 114505}.

\bibitem{Cheng:2020qzc}
X. D. Cheng and R. M. Wang and Y. G. Xu,
Study of $a_{0}^{0} \left( 980 \right)-f_{0} \left( 980 \right)$ mixing from $a_{0}(1450) \to a_{0}^{0}(980) f_{0}(500) \to \pi^{+} \pi^{-} f_{0}(500)$,
\href{https://doi.org/10.1103/PhysRevD.102.054009}
{Phys. Rev. D \textbf{102} (2020) 054009}.
[\href{https://arxiv.org/pdf/2007.15210}
{arXiv:2007.15210}]

\bibitem{Lee:1999kv}
W.~J.~Lee and D.~Weingarten,
Scalar quarkonium masses and mixing with the lightest scalar glueball,
\href{https://doi.org/10.1103/PhysRevD.61.014015}
{Phys. Rev. D \textbf{61} (2000) 014015}.
[\href{https://arxiv.org/pdf/hep-lat/9910008}
{hep-lat/9910008}]


\bibitem{ParticleDataGroup:2024cfk}
S.~Navas \textit{et al}. [Particle Data Group],
Review of particle physics,
\href{https://doi.org/10.1103/PhysRevD.110.030001}
{Phys. Rev. D \textbf{110} (2024) 030001}.


\bibitem{Cason:1976fn}
N.~M.~Cason, V.~A.~Polychronakos, J.~M.~Bishop, N.~N.~Biswas, V.~P.~Kenney, D.~S.~Rhines, W.~D.~Shephard and J.~M.~Watson,
Observation of a New Scalar Meson,
\href{https://doi.org/10.1103/PhysRevLett.36.1485}
{Phys. Rev. Lett. \textbf{36} (1976) 1485}.


\bibitem{OBELIX:1998sco}
A.~Bertin \textit{et al}. [OBELIX],
Study of the isovector scalar mesons in the channel $p\bar{p}\to K^{\pm} K^0_S \pi^{\mp}$ at rest with initial angular momentum state selection,
\href{https://doi.org/10.1016/S0370-2693(98)00767-9}
{Phys. Lett. B \textbf{434} (1998) 180-188}.

\bibitem{Belle:2009xpa}
S.~Uehara \textit{et al}. [Belle],
High-statistics study of $\eta \pi^0$ production in two-photon collisions,
\href{https://doi.org/10.1103/PhysRevD.80.032001}
{Phys. Rev. D \textbf{80} (2009) 032001}.
[\href{https://arxiv.org/pdf/0906.1464}
{arXiv:0906.1464}]


\bibitem{Liu:2019tsi}
K.~Liu [BESIII Collaboration],
Semileptonic and leptonic $D$ decays at BESIII,
\href{https://doi.org/10.22323/1.367.0046}
{PoS \textbf{LeptonPhoton2019} (2019) 046}.

\bibitem{Zhang:2019tcs}
S.~Zhang [BESIII Collaboration],
Test lepton flavor universality with (semi)leptonic $D$ decays at BESIII,
\href{https://scipost.org/10.21468/SciPostPhysProc.1.016}
{SciPost Phys. Proc. \textbf{1} (2019) 016}.

\bibitem{Yang:2018qdx}
Y.~H.~Yang [BESIII Collaboration],
(Semi-)leptonic decays of $D$ Mesons at BESIII,
\href{https://arxiv.org/abs/1812.00320}
{arXiv:1812.00320}.

\bibitem{BESIII:2016gbw}
M.~Ablikim \textit{et al}. [BESIII Collaboration],
Improved measurement of the absolute branching fraction of $D^{+}\to \bar{K}^0 \mu ^{+}\nu _{\mu }$,
\href{https://doi.org/10.1140/epjc/s10052-016-4198-2}
{Eur. Phys. J. C \textbf{76} (2016) 369}.
[\href{https://arxiv.org/abs/1605.00068}
{arXiv:1605.00068}]

\bibitem{BESIII:2021mfl}
M.~Ablikim \textit{et al}. [BESIII Collaboration],
Determination of the absolute branching fractions of $D^0\to K^-e^+\nu_e$ and $D^+\to \bar K^0 e^+\nu_e$,
\href{https://doi.org/10.1103/PhysRevD.104.052008}
{Phys. Rev. D \textbf{104} (2021) 052008}.
[\href{https://arxiv.org/abs/2104.08081}
{arXiv:2104.08081}]

\bibitem{BESIII:2015tql}
M.~Ablikim \textit{et al}. [BESIII Collaboration],
Study of Dynamics of $D^0 \to K^- e^+ \nu_{e}$ and $D^0\to\pi^- e^+ \nu_{e}$ Decays,
\href{https://doi.org/10.1103/PhysRevD.92.072012}
{Phys. Rev. D \textbf{92} (2015) 072012}.
[\href{https://arxiv.org/abs/1508.07560}
{arXiv:1508.07560}]

\bibitem{BESIII:2021pvy}
M.~Ablikim \textit{et al}. [BESIII Collaboration],
Observation of the decay $D^0\to \rho^-\mu^+\nu_\mu$,
\href{https://doi.org/10.1103/PhysRevD.104.L091103}
{Phys. Rev. D \textbf{104} (2021) L091103}.
[\href{https://arxiv.org/abs/2106.02292}
{arXiv:2106.02292}]

\bibitem{BESIII:2015kin}
M.~Ablikim \textit{et al}. [BESIII Collaboration],
Measurement of the form factors in the decay $D^+ \to \omega e^+ \nu_{e}$ and search for the decay $D^+ \to \phi e^+ \nu_{e}$,
\href{https://doi.org/10.1103/PhysRevD.92.071101}
{Phys. Rev. D \textbf{92} (2015) 071101}.
[\href{https://arxiv.org/abs/1508.00151}
{arXiv:1508.00151}]

\bibitem{BaBar:2014xzf}
J.~P.~Lees \textit{et al}. [BaBar Colleboration],
Measurement of the $D^0 \to \pi^- e^+ \nu_e$ differential decay branching fraction as a function of $q^2$ and study of form factor parameterizations,
\href{https://doi.org/10.1103/PhysRevD.91.052022}
{Phys. Rev. D \textbf{91} (2015) 052022}.
[\href{/https://arxiv.org/pdf/1412.5502}
{arXiv:1412.5502}]

\bibitem{Belle:2006idb}
L.~Widhalm \textit{et al}. [Belle Colleboration],
Measurement of $D^0 \to \pi \ell \nu (K \ell \nu)$ Form Factors and Absolute branching fractions,
\href{https://doi.org/10.1103/PhysRevLett.97.061804}
{Phys. Rev. Lett. \textbf{97} (2006) 061804}.
[\href{https://arxiv.org/pdf/hep-ex/0604049}
{arXiv:0604049}]

\bibitem{CLEO:2011ab}
S.~Dobbs \textit{et al}. [CLEO Collaboration],
First Measurement of the Form Factors in the Decays $D^0 \to \rho^- e^+ \nu_e$ and $D^+ \to \rho^0 e^+ \nu_e$,
\href{https://doi.org/10.1103/PhysRevLett.110.131802}
{Phys. Rev. Lett. \textbf{110} (2013) 131802}.
[\href{https://arxiv.org/abs/1112.2884}
{arXiv:1112.2884}]

\bibitem{CLEO:2004arv}
G.~S.~Huang \textit{et al}. [CLEO Collaboration],
Study of semileptonic charm decays $D^0\to \pi^- \ell^+ \nu$ and $D^0\to K^- \ell^+\nu$,
\href{https://doi.org/10.1103/PhysRevLett.94.011802}
{Phys. Rev. Lett. \textbf{94} (2005) 011802}.
[\href{https://arxiv.org/abs/hep-ex/0407035}
{arXiv:0407035}]

\bibitem{CLEO:2009dyb}
J.~Yelton \textit{et al}. [CLEO Collaboration],
Absolute branching Fraction Measurements for Exclusive $D_s$ Semileptonic Decays,
\href{https://doi.org/10.1103/PhysRevD.80.052007}
{Phys. Rev. D \textbf{80} (2009) 052007}.
[\href{https://arxiv.org/abs/0903.0601}
{arXiv:0903.0601}]

\bibitem{CLEO:2009svp}
D.~Besson \textit{et al}. [CLEO Collaboration],
Improved measurements of $D$ meson semileptonic decays to $\pi$ and $K$ mesons,
\href{https://doi.org/10.1103/PhysRevD.80.032005}
{Phys. Rev. D \textbf{80} (2009) 032005}.
[\href{https://arxiv.org/abs/0906.2983}
{arXiv:0906.2983}]

\bibitem{CLEO:2005rxg}
G.~S.~Huang \textit{et al}. [CLEO Collaboration],
Absolute branching fraction measurements of exclusive $D^+$ semileptonic decays,
\href{https://doi.org/10.1103/PhysRevLett.95.181801}
{Phys. Rev. Lett. \textbf{95} (2005) 181801}.
[\href{https://arxiv.org/abs/hep-ex/0506053}
{arXiv:0506053}]

\bibitem{Verma:2011yw}
R.~C.~Verma,
Decay constants and form factors of s-wave and p-wave mesons in the covariant light-front quark model,
\href{https://doi:10.1088/0954-3899/39/2/025005}
{J. Phys. G \textbf{39} (2012) 025005}.
[\href{https://arxiv.org/pdf/1103.2973}
{arXiv:1103.2973}]

\bibitem{Galkin:2025emi}
V.~O.~Galkin and I.~S.~Sukhanov,
Exclusive semileptonic decays of $D$ and $D_s$ mesons into orbitally and radially excited states of strange and light mesons,
\href{https://doi:10.1103/PhysRevD.111.093001}
{Phys. Rev. D \textbf{111} (2025) 093001}.
[\href{https://arxiv.org/pdf/2501.16406}
{arXiv:2501.16406}]

\bibitem{Huang:2021owr}
Q.~Huang, Y.~J.~Sun, D.~Gao, G.~H.~Zhao, B.~Wang and W.~Hong,
Study of form factors and branching ratios for $D\rightarrow S,Al\bar{\nu_{l}}$ with light-cone sum rules,
\href{https://arxiv.org/pdf/2102.12241}.
{arXiv:2102.12241}.

\bibitem{Balitsky:1989ry}
I.~I.~Balitsky, V.~M.~Braun and A.~V.~Kolesnichenko,
Radiative Decay $\sigma^+ \to p \gamma$ in Quantum Chromodynamics,
\href{https://doi:10.1016/0550-3213(89)90570-1}
{Nucl. Phys. B \textbf{312} (1989) 509-550}.

\bibitem{Chernyak:1990ag}
V.~L.~Chernyak and I.~R.~Zhitnitsky,
B meson exclusive decays into baryons,
\href{https://doi:10.1016/0550-3213(90)90612-H}
{Nucl. Phys. B \textbf{345} (1990) 137-172}.

\bibitem{Cheng:2017bzz}
W.~Cheng, X.~G.~Wu and H.~B.~Fu,
Reconsideration of the $B \to K^*$ transition form factors within the QCD light-cone sum rules,
\href{https://doi:10.1103/PhysRevD.95.094023}
{Phys. Rev. D \textbf{95} (2017) 094023}.
[\href{https://arxiv.org/pdf/1703.08677}
{arXiv:1703.08677}]

\bibitem{Duplancic:2008ix}
G.~Duplancic, A.~Khodjamirian, T.~Mannel, B.~Melic and N.~Offen,
Light-cone sum rules for $B\to \pi$ form factors revisited,
\href{https://doi.org/10.1088/1126-6708/2008/04/014}
{JHEP \textbf{04} (2008), 014}.
[\href{https://arxiv.org/abs/0801.1796}
{arXiv:0801.1796}]

\bibitem{Tian:2023vbh}
H.~J.~Tian, H.~B.~Fu, T.~Zhong, X.~Luo, D.~D.~Hu and Y.~L.~Yang,
Investigating the $D_s^{+} \to \pi^{0} \ell^{+} \nu_\ell $ decay process within the QCD sum rule approach,
\href{https://doi:10.1103/PhysRevD.108.076003}
{Phys. Rev. D \textbf{108} (2023) 076003}.
[\href{https://arxiv.org/pdf/2306.07595}
{arXiv:2306.07595}]

\bibitem{Gao:2019lta}
J.~Gao, C.~D.~L\"u, Y.~L.~Shen, Y.~M.~Wang and Y.~B.~Wei,
Precision calculations of $B \to V$ form factors from soft-collinear effective theory sum rules on the light-cone,
\href{https://doi:10.1103/PhysRevD.101.074035}
{Phys. Rev. D \textbf{101} (2020) 074035}.
[\href{https://arxiv.org/pdf/1907.11092}
{arXiv:1907.11092}]

\bibitem{LatticeParton:2022zqc}
J.~Hua \textit{et al}. [Lattice Parton],
Pion and Kaon Distribution Amplitudes from Lattice QCD,
\href{https://doi:10.1103/PhysRevLett.129.132001}
{Phys. Rev. Lett. \textbf{129} (2022) 132001}.
[\href{https://arxiv.org/pdf/2201.09173}
{arXiv:2201.09173}]

\bibitem{Huang:1994dy}
T.~Huang, B.~Q.~Ma and Q.~X.~Shen,
Analysis of the pion wave function in light cone formalism,
\href{https://doi.org/10.1103/PhysRevD.49.1490}
{Phys. Rev. D \textbf{49} (1994) 1490-1499}.
[\href{https://arxiv.org/pdf/hep-ph/9402285}
{arXiv:9402285}]

\bibitem{Dassinger:2008as}
B.~Dassinger, R.~Feger and T.~Mannel,
Complete Michel Parameter Analysis of inclusive semileptonic $b\to c$ transition,
\href{https://doi.org/10.1103/PhysRevD.79.075015}
{Phys. Rev. D \textbf{79} (2009) 075015}.
[\href{https://arxiv.org/pdf/0803.3561}
{arXiv:0803.3561}]


\bibitem{Fu:2013wqa}
H.~B.~Fu, X.~G.~Wu, H.~Y.~Han, Y.~Ma and T.~Zhong,
$|V_{cb}|$ from the semileptonic decay $B\to D \ell \bar{\nu}_\ell $ and the properties of the $D$ meson distribution amplitude,
\href{https://doi.org/10.1016/j.nuclphysb.2014.04.021}
{Nucl. Phys. B \textbf{884} (2014) 172-192}.
[\href{https://arxiv.org/pdf/1309.5723}
{arXiv:1309.5723}]

\bibitem{Cheng:2017fkw}
X.~D.~Cheng, H.~B.~Li, B.~Wei, Y.~G.~Xu and M.~Z.~Yang,
Study of $D \to a_0 (980) e^+ \nu_e$ decay in the light-cone sum rules approach,
\href{https://doi.org/10.1103/PhysRevD.96.033002}
{Phys. Rev. D \textbf{96} (2017) 033002}.
[\href{/https://arxiv.org/pdf/1706.01019}
{arXiv:1706.01019}]

\bibitem{Cui:2022zwm}
B.~Y.~Cui, Y.~K.~Huang, Y.~L.~Shen, C.~Wang and Y.~M.~Wang,
Precision calculations of $B_{d,s} \to \pi,K$ decay form factors in soft-collinear effective theory,
\href{https://doi.org/10.1007/JHEP03(2023)140}
{JHEP \textbf{03} (2023) 140}.
[\href{https://arxiv.org/abs/2212.11624}
{arXiv:2212.11624}]

\bibitem{Wu:2011gf}
X.~G.~Wu and T.~Huang,
Constraints on the Light Pseudoscalar Meson Distribution Amplitudes from Their Meson-Photon Transition Form Factors,
\href{https://doi.org/10.1103/PhysRevD.84.074011}
{Phys. Rev. D \textbf{84} (2011) 074011}.
[\href{https://arxiv.org/pdf/1106.4365}
{arXiv:1106.4365}]

\bibitem{Wu:2010zc}
X.~G.~Wu and T.~Huang,
An Implication on the Pion Distribution Amplitude from the Pion-Photon Transition Form Factor with the New BABAR Data,
\href{https://doi.org/10.1103/PhysRevD.82.034024}
{Phys. Rev. D \textbf{82} (2010) 034024}.
[\href{https://arxiv.org/pdf/1005.3359}
{arXiv:1005.3359}]

\bibitem{Huang:2004su}
T.~Huang and X.~G.~Wu,
A Model for the twist-3 wave function of the pion and its contribution to the pion form-factor,
\href{https://doi.org/10.1103/PhysRevD.70.093013}
{Phys. Rev. D \textbf{70} (2004) 093013}.
[\href{https://arxiv.org/pdf/hep-ph/0408252}
{hep-ph/0408252}]

\bibitem{Cao:1997hw}
F.~g.~Cao and T.~Huang,
Large corrections to asymptotic F(eta c gamma) and F(eta b gamma) in the light cone perturbative QCD,
\href{https://doi.org/10.1103/PhysRevD.59.093004}
{Phys. Rev. D \textbf{59} (1999) 093004}.
[\href{https://arxiv.org/pdf/hep-ph/9711284}
{hep-ph/9711284}]


\bibitem{Zhong:2021epq}
T.~Zhong, Z.~H.~Zhu, H.~B.~Fu, X.~G.~Wu and T.~Huang,
Improved light-cone harmonic oscillator model for the pionic leading-twist distribution amplitude,
\href{https://doi:10.1103/PhysRevD.104.016021}
{Phys. Rev. D \textbf{104} (2021) 016021}.
[\href{https://arxiv.org/pdf/2102.03989}
{arXiv:2102.03989}]

\bibitem{Wu:2007rt}
X.~G.~Wu, T.~Huang and Z.~Y.~Fang,
$B\to K$ transition form-factor up to $\mathcal{O}(1/m_b^2)$ within the $k_T$ factorization approach,
\href{https://doi.org/10.1140/epjc/s10052-007-0421-5}
{Eur. Phys. J. C \textbf{52} (2007) 561-570}.
[\href{/https://arxiv.org/pdf/0707.2504}
{arXiv:0707.2504}]

\bibitem{Ma:1993ht}
B.~Q.~Ma,
Spin structure of the pion in a light cone representation,
\href{https://doi.org/10.1007/BF01280840}
{Z. Phys. A \textbf{345} (1993) 321-325}.
[\href{https://arxiv.org/pdf/hep-ph/9305283}
{hep-ph/9305283}]

\bibitem{Lepage:1980fj}
G.~P.~Lepage and S.~J.~Brodsky,
Exclusive Processes in Perturbative Quantum Chromodynamics,
\href{https://doi.org/10.1103/PhysRevD.22.2157}
{Phys. Rev. D \textbf{22} (1980) 2157}.

\bibitem{Fu:2014uea}
H.~B.~Fu, X.~G.~Wu and Y.~Ma,
$B\to K^*$ Transition Form Factors and the Semi-leptonic Decay $B \to K^* \mu^+ \mu^-$,
\href{https://iopscience.iop.org/article/10.1088/0954-3899/43/1/015002}
{J. Phys. G \textbf{43} (2016) 015002}.
[\href{https://arxiv.org/pdf/1411.6423}
{arXiv:1411.6423}]

\bibitem{Zhong:2011rg}
T.~Zhong, X.~G.~Wu, H.~Y.~Han, Q.~L.~Liao, H.~B.~Fu and Z.~Y.~Fang,
Revisiting the Twist-3 Distribution Amplitudes of $K$ Meson within the QCD Background Field Approach,
\href{https://doi.org/10.1088/0253-6102/58/2/16}
{Commun. Theor. Phys. \textbf{58} (2012) 261-270}.
[\href{https://arxiv.org/pdf/1109.3127}
{arXiv:1109.3127}]


\bibitem{Lu:2006fr}
C.~D.~Lu, Y.~M.~Wang and H.~Zou,
Twist-3 distribution amplitudes of scalar mesons from QCD sum rules,
\href{https://doi.org/10.1103/PhysRevD.75.056001}
{Phys. Rev. D \textbf{75} (2007) 056001}.
[\href{https://arxiv.org/pdf/hep-ph/0612210}
{hep-ph/0612210}]

\bibitem{ParticleDataGroup:2016lqr}
C.~Patrignani \textit{et al}. [Particle Data Group],
Review of Particle Physics,
\href{https://iopscience.iop.org/article/10.1088/1674-1137/40/10/100001}
{Chin. Phys. C \textbf{40} (2016) 100001}.

\bibitem{Ball:2006nr}
P.~Ball and R.~Zwicky,
$|V_{td} / V_{ts}|$ from $B \to V \gamma$,
\href{https://iopscience.iop.org/article/10.1088/1126-6708/2006/04/046}
{JHEP \textbf{04} (2006) 046}.
[\href{https://arxiv.org/pdf/hep-ph/0603232}
{hep-ph/0603232}]

\bibitem{Ball:2004ye}
P.~Ball and R.~Zwicky,
New results on $B \to \pi, K, \eta$ decay formfactors from light-cone sum rules,
\href{https://doi.org/10.1103/PhysRevD.71.014015}
{Phys. Rev. D \textbf{71} (2005) 014015}.
[\href{/https://arxiv.org/pdf/hep-ph/0406232}
{hep-ph/0406232}]

\bibitem{Fu:2018yin}
H.~B.~Fu, L.~Zeng, R.~L\"u, W.~Cheng and X.~G.~Wu,
The $D\to \rho$ semileptonic and radiative decays within the light-cone sum rules,
\href{https://doi.org/10.1140/epjc/s10052-020-7758-4}
{Eur. Phys. J. C \textbf{80} (2020) 194}.
[\href{https://arxiv.org/pdf/1808.06412}
{arXiv:1808.06412}]

\bibitem{Cheng:2017pcq}
H.~Y.~Cheng and X.~W.~Kang,
Branching fractions of semileptonic $D$ and $D_s$ decays from the covariant light-front quark model,
\href{https://doi.org/10.1140/epjc/s10052-017-5170-5}
{Eur. Phys. J. C \textbf{77} (2017) 587}.
[\href{https://arxiv.org/pdf/1707.02851}
{arXiv:1707.02851}]











































\end{thebibliography}
\end{document}